\documentclass[10pt,aps,prd,nofootinbib,floatfix,twocolumn,superscriptaddress,groupedaddress]{revtex4-2}
\usepackage{ragged2e}
\usepackage{amsmath,amssymb}
\usepackage{bbm}
\usepackage{mathrsfs}
\usepackage{graphicx}
\usepackage{float}
\usepackage{caption}
\usepackage{subcaption}
\usepackage{slashed}
\usepackage{color,soul}
\usepackage[dvipsnames]{xcolor}
\usepackage{mathtools}
\usepackage{dsfont}
\usepackage{hyperref}
\usepackage[noabbrev]{cleveref}
\hypersetup{colorlinks=true, linkcolor=blue, citecolor=blue}
\usepackage{orcidlink}
\usepackage{stackrel}

\begin{document}

\makeatletter
\newbox\slashbox \setbox\slashbox=\hbox{$/$}
\newbox\Slashbox \setbox\Slashbox=\hbox{\large$/$}
\def\pFMslash#1{\setbox\@tempboxa=\hbox{$#1$}
  \@tempdima=0.5\wd\slashbox \advance\@tempdima 0.5\wd\@tempboxa
  \copy\slashbox \kern-\@tempdima \box\@tempboxa}
\def\pFMSlash#1{\setbox\@tempboxa=\hbox{$#1$}
  \@tempdima=0.5\wd\Slashbox \advance\@tempdima 0.5\wd\@tempboxa
  \copy\Slashbox \kern-\@tempdima \box\@tempboxa}
\def\FMslash{\protect\pFMslash}
\def\FMSlash{\protect\pFMSlash}
\def\miss#1{\ifmmode{/\mkern-11mu #1}\else{${/\mkern-11mu #1}$}\fi}
\makeatother

\title{New physics in the $ZZh$ vertex: One-loop contributions from a radiative seesaw model }

\author{H\'ector Novales-S\'anchez$^{(a)}$}
\author{Humberto V\'azquez-Castro$^{(a)}$}
\author{M\'onica Salinas$^{(b)}$}
\affiliation{
$^{(a)}$Facultad de Ciencias F\'isico Matem\'aticas, Benem\'erita Universidad Aut\'onoma de Puebla, Apartado Postal 1152 Puebla, Puebla, M\'exico\\$^{(b)}$Departamento de F\'isica, Centro de Investigaci\'on y de Estudios Avanzados del IPN, Apartado Postal 14-740, 07000 Ciudad de M\'exico, M\'exico}

\begin{abstract}
Precision Higgs physics offers a sensitive window into physics beyond the Standard Model. In parallel, neutrino-oscillation experiments have established the existence of nonzero neutrino masses, thus implying the presence of new physics. Motivated by these facts, we investigate the one-loop contributions of light and heavy Majorana neutrinos to the $ZZh$ vertex within a variant of the type-I seesaw mechanism in which light-neutrino masses vanish at tree level and are then generated radiatively. We analyze the $CP$-conserving and $CP$-violating anomalous couplings which characterize the $ZZh$ vertex and study their phenomenological implications in two relevant kinematic scenarios at future lepton colliders: Higgsstrahlung production and Higgs production via neutral-current vector-boson fusion. We find that $CP$-conserving effects can reach magnitudes of order $10^{-3}$, potentially within projected future experimental sensitivities, whereas $CP$-violating contributions are strongly suppressed, lying well beyond such projections.
\end{abstract}

\pacs{.}
\maketitle

\section{Introduction}
A main event for the high-energy-physics community has been doubtless the measurement, by the ATLAS and CMS collaborations at the Large Hadron Collider (LHC), of a scalar Higgs-like particle, with mass $~\sim125\,\textrm{GeV}$~\cite{Aad_2012,Chatrchyan_2012}. However, before finally arriving at the conclusion that such a particle actually corresponds to the one given by the minimal scalar sector characterizing the Standard Model (SM), a huge amount of work, on both the experimental and theoretical sides, remains to be done. The Higgs couplings to the electroweak gauge bosons play a particularly relevant role, as they are directly linked to the mechanism of electroweak symmetry breaking and are highly sensitive to possible new-physics effects. In this context, the $ZZh$ vertex, which describes the interaction between two $Z$ gauge bosons and the Higgs boson, provides a particularly clean probe of the gauge structure of the scalar sector, as its form is completely determined at tree level by gauge invariance~\cite{Djouadi_2008}. In the SM, this coupling exhibits a well-defined Lorentz structure given by
\begin{equation}
\begin{gathered}
            \includegraphics[scale=1]{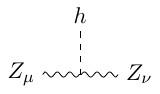}
    \end{gathered}=i\,\frac{g}{c_\textrm{w}}\,m_Z\,g_{\mu\nu},
    \label{ZZhSM}
\end{equation}
where $g$ denotes the $\textrm{SU}(2)_L$ gauge coupling, $m_Z$ is the $Z$-boson mass, and $c_\textrm{w}=\cos\theta_\textrm{w}$, with $\theta_\textrm{w}$ the weak mixing angle.\\

From a phenomenological perspective, the $ZZh$ vertex participates in some of the most important processes for Higgs studies. For instance, the decay $h\to 4\ell$ is regarded as the golden channel, having played a crucial role in the Higgs discovery~\cite{Aad_2012,Chatrchyan_2012} and in subsequent measurements of its mass, spin, and properties~\cite{Chatrchyan_2014,Aad_2014,2016}.
This coupling also contributes to production mechanisms such as neutral-current vector-boson fusion (VBF) and, notably, associated production $pp\to Zh$ at hadron colliders, which allows for detailed studies of decay modes like $h\to b\overline{b}$~\cite{Aad:2915296,ATLAS-CONF-2017-055,2021}. Another particularly interesting scenario is off-shell Higgs production in the process $gg\to ZZ$, which enables indirect access to the total Higgs decay width~\cite{CMS2022,ATLAS2015,Aad_2016}.\\

At future lepton colliders, such as the Compact Linear Collider (CLIC)~\cite{CLIC} and the International Linear Collider (ILC)~\cite{ILC}, the Higgsstrahlung process $e^{-}e^{+}\to Zh$ will allow for precision measurements of the $ZZh$ coupling, both in its magnitude and in its Lorentz structure~\cite{ILCHiggs,Borzumati_2014,Tian:2013yda}. In this clean experimental environment, subtle deviations from the SM prediction can be probed with high sensitivity.\\

One-loop corrections to the $ZZh$ vertex have been computed within the SM framework~\cite{PhysRevD.23.2001,KNIEHL19911,Phan:2022amy}, and $CP$-violating effects have also been investigated~\cite{PhysRevD.107.115031}. Furthermore, new-physics contributions to the $ZZh$ interaction have been reported in several extensions of the SM, including the two-Higgs-doublet model~\cite{PhysRevD.108.095013,KIKUCHI2016807}, the Higgs singlet model~\cite{KANEMURA2016286,PhysRevD.96.035014}, supersymmetric models~\cite{Baglio_2020,Englert:2014ffa}, the inert Higgs doublet model~\cite{Arhrib_2015,PhysRevD.94.115011}, the minimal Higgs triplet model~\cite{PhysRevD.87.015012}, and multi-Higgs models~\cite{Chiang_2017,Grossman_1994}.\\

With this motivation, in the present work we consider the variant of the seesaw mechanism introduced in Ref.~\cite{Pilaftsis}, in which the light-neutrino masses vanish at tree level. In this framework, the light-neutrino masses are generated radiatively through two-point correlation functions, while the heavy-neutrino mass spectrum becomes quasi-degenerate. The one-loop contributions to the $ZZh$ vertex arise from the couplings of neutrinos to the Higgs boson and to the $Z$ boson. The corresponding Lagrangians are given by
\begin{eqnarray}
&&
\mathcal{L}_{hnn}=\frac{g}{4m_W}h\,\overline{n}\Big( P_R(M_n\mathcal{C}^*+\mathcal{C}M_n)
\nonumber \\ && \hspace{1cm}
+P_L(M_n\mathcal{C}+\mathcal{C}^*M_n) \Big)n,
\label{Lhnn}
\end{eqnarray}
\begin{equation}
\mathcal{L}_{Znn}=\frac{g}{4c_W}Z_\mu\overline{n}\gamma^\mu\big( \mathcal{C}P_L-\mathcal{C}^*P_R \big)n.
\label{LNC}
\end{equation}

In order to write these expressions, we have adopted the notation of Refs.~\cite{WWA,ZZZ,WWZ,CLFV,WWh}, in which light and heavy neutrino fields have been collectively denoted by $n_j$, with $n_1=\nu_1$, $n_2=\nu_2$, $n_3=\nu_3$, $n_4=N_1$, $n_5=N_2$, and $n_6=N_3$. Here, $\nu_j$ and $N_j$, with $j=1,2,3$, denote the light and heavy neutrino fields, with corresponding masses $m_{\nu_j}$ and $m_{N_j}$. Both light and heavy neutrino fields are Majorana fields and therefore satisfy the Majorana condition $n_j^{\rm c}=n_j$. We further define the $6\times1$ column vector $n$ by its components $(n)_j=n_j$, with $j=1,\dots,6$. Eq.~\eqref{Lhnn} involves the diagonal neutrino-mass matrix $M_n$, whose entries are given by $\big(M_n\big)_{jk}=\delta_{jk}m_{n_k}$. Furthermore, Eqs.~\eqref{Lhnn} and \eqref{LNC} include the $6\times6$ Hermitian matrix $\mathcal{C}$, which describes the mixing structure of light and heavy Majorana neutrinos and can be written in terms of $3\times3$ blocks as
\begin{equation}
{\cal C}=
\left(
\begin{array}{cc}
{\cal C}_{\nu\nu} & {\cal C}_{\nu N}
\vspace{0.3cm}
\\
{\cal C}_{N\nu} & {\cal C}_{NN}
\end{array}
\right).
\label{Cmatrix}
\end{equation}

The matrix $\mathcal{C}$ determines the couplings of neutrinos to the Higgs and $Z$ bosons in the mass basis. The submatrices ${\cal C}_{\nu\nu}$, ${\cal C}_{\nu N}$, ${\cal C}_{N\nu}$, and ${\cal C}_{NN}$ describe light-neutrino interactions, light--heavy and heavy--light transitions, and heavy-neutrino interactions, respectively. The matrix ${\cal C}$ further satisfies the relation
${\cal C}{\cal C}^\dagger = {\cal C}$.
Additional theoretical details of the model can be found in Refs.~\cite{WWA,ZZZ,WWZ}.\\

We investigate the one-loop contributions of light and heavy Majorana neutrinos to the $ZZh$ vertex. Our results indicate that $CP$-conserving effects can reach values of order $10^{-3}$, potentially comparable to the projected sensitivities of future lepton colliders, whereas $CP$-violating contributions, found to be quite suppressed, turn out to be of order $10^{-15}$.\\

This paper is organized as follows. In Section~\ref{ZZhvertex}, we introduce the Lorentz-covariant parametrization of the $ZZh$ vertex and then present the analytic one-loop calculation of new-physics contributions to the anomalous couplings (ACs) that characterize such an interaction. In Section~\ref{estimations}, we present our numerical analysis and discussion on the ACs. Our conclusions are presented in Section~\ref{summary}.

\section{The $ZZh$ vertex and new physics contributions}
\label{ZZhvertex}
In this section, we introduce the parametrization employed for the phenomenological analysis of the $ZZh$ vertex within the framework of the Strongly-Interacting Light Higgs (SILH) effective-Lagrangian basis. We then present the analytical calculation of the one-loop contributions arising from Majorana neutrinos.

\subsection{Anomalous $ZZh$ couplings}

The effective field theory formalism offers a systematic description of new-physics effects associated with high-energy scales, with the effective Lagrangian constructed from the low-energy fields and symmetries~\cite{WUDKA,Dobado:1997jx,Carsten}. In particular, the effective Lagrangian for the SM, $\mathcal{L}_{\rm SM}^{\rm eff}$,  has the general form
\begin{equation}
    \mathcal{L}_{\rm SM}^{\rm eff}=\mathcal{L}_{\rm SM}+\sum_{n>4}^{\infty}\sum_{j}^{j_n}\frac{\alpha_{j}^{(n)}}{\Lambda^{n-4}}\mathcal{O}_{j}^{(n)}.
\end{equation}
This Lagrangian is governed by the gauge-symmetry group $\rm SU(3)_C\otimes\rm SU(2)_L\otimes\rm U(1)_Y$, and is given by the usual SM Lagrangian, $\mathcal{L}_{\text{SM}}$, followed by an infinite sum of operators $\mathcal{O}_{j}^{(n)}$ of canonical dimension $n>4$, which parametrize, at current achievable experimental sensitivity, the low-energy effects of unknown new physics characterized by a high-energy scale $\Lambda$. The coefficients $\alpha_{j}^{(n)}$ are dimensionless quantities, commonly referred to as ``Wilson coefficients''. The factors $\frac{1}{\Lambda^{n-4}}$ ensure that each term in $\mathcal{L}_{\rm SM}^{\rm eff}$ has overall units $(\textrm{mass})^4$. Non-renormalizable terms of mass dimensions $n=6,8,10,\cdots$ are the only ones appearing in the infinite series as long as lepton number is a conserved quantity. Otherwise, effective-Lagrangian operators with odd mass dimensions $n=5,7,9,\cdots$~\cite{Babu_2001} are allowed. For instance, if lepton-number violation is assumed, the only dimension-five operator that can be constructed solely from SM fields is the Weinberg operator~\cite{Weinberg}, which generates Majorana neutrino masses after electroweak symmetry breaking.\\

One of the main appeals of effective Lagrangians is the study of deviations from the SM predictions, which translate directly into bounds on the Wilson coefficients~\cite{Leung:1984ni,BUCHMULLER1986621}. In particular, corrections to the $ZZh$ vertex arise from the set of mass-dimension-6 effective operators in the Lagrangian $\mathcal{L}_{\rm SM}^{\rm eff}$, which modify both the strength of the coupling and its Lorentz-covariant structure, including possible $CP$-violating contributions.\\

Gauge and Lorentz symmetry criteria allow for the construction of a large number of effective terms; however, not all of them are independent, since relations emerge from the use of equations of motion~\cite{WUDKA,Carsten}, integration by parts, Bianchi identities, and Fierz identities. Therefore, the new-physics encoded in an effective-field theory can be actually described within different bases of effective operators $\mathcal{O}_{j}^{(n)}$. In particular, this is the case of the subset of effective-Lagrangian terms comprised by Higgs-associated effective operators, for which the Warsaw basis~\cite{Grzadkowski_2010} and the Higgs boson basis~\cite{LHCHiggs,Azatov:2022kbs} are two options. The relations among the Wilson coefficients of these bases have been discussed, for instance, in Refs.~\cite{Marzocca2020BSMBF,Falkowski:2015wza,Falkowski:2001958,CAO2025116781}. Another convenient parametrization is provided by the SILH basis, originally introduced to describe scenarios in which the Higgs boson emerges as a light state from a strongly interacting sector characterized by a scale $\Lambda$. The SILH Lagrangian is given by~\cite{SILH,SILH1,SILH2}
\begin{widetext}
\begin{multline}
    \mathcal{L}_{\rm SILH}=\frac{\overline{c}_\Phi}{2v^2}\partial^\mu\big( \Phi^{\dagger}\Phi \big)\partial_\mu\big( \Phi^{\dagger}\Phi \big)+\frac{\overline{c}_{T}}{2v²}\big(\Phi^{\dagger}\overset{\text{\scriptsize$\leftrightarrow$}}{D}\hspace{0.0000000001cm}^\mu\Phi \big)\big(\Phi^{\dagger}\overset{\text{\scriptsize$\leftrightarrow$}}{D}\hspace{0.0000000001cm}_\mu\Phi \big)+\frac{ig\overline{c}_W}{2m_W^2}\big(\Phi^{\dagger} \sigma^j \overset{\text{\scriptsize$\leftrightarrow$}}{D}\hspace{0.0000000001cm}^\mu\Phi \big) D^\nu W^j_{\mu\nu}\\+\frac{ig^{'}\overline{c}_{B}}{2m_W^{2}}\big(\Phi^{\dagger}\overset{\text{\scriptsize$\leftrightarrow$}}{D}\hspace{0.0000000001cm}^\mu\Phi \big)\partial^{\nu}B_{\mu\nu}+\frac{ig\overline{c}_{\Phi W}}{m_W^2}\big( (D^\mu\Phi)^{\dagger}\sigma^jD^\nu\Phi \big)W^j_{\mu\nu}+\frac{ig^{'}\overline{c}_{\Phi B}}{m_W^{2}}\big(D^{\mu}\Phi\big)^{\dagger}\big(D^{\nu}\Phi\big)B_{\mu\nu}\\+\frac{g^{' 2}\overline{c}_{\gamma}}{m_W^{2}}\Phi^{\dagger}\Phi B_{\mu\nu}B^{\mu\nu}+\frac{ig\tilde{c}_{\Phi W}}{m_W^2}\big( (D^\mu\Phi)^{\dagger}\sigma^jD^\nu\Phi \big)\tilde{W}^j_{\mu\nu}+\frac{ig^{'}\tilde{c}_{\Phi B}}{m_W^{2}}\big(D^{\mu}\Phi\big)^{\dagger}\big(D^{\nu}\Phi\big)B_{\mu\nu}\\+\frac{g^{' 2}\tilde{c}_{\gamma}}{m_W^{2}}\Phi^{\dagger}\Phi B_{\mu\nu}\tilde{B}^{\mu\nu}+\cdots,
    \label{LSILH}
\end{multline}
\end{widetext}
where $\Phi$ is the SM Higgs doublet, $D_\mu$ is the $\rm SU(2)_L\otimes\rm U(1)_Y$ covariant derivative, $W_{\mu\nu}^{j}=\partial_\mu W_\nu^{j}-\partial_\nu W_{\mu}^{j}-g\hspace{0.5mm}\epsilon^{jki}W_\mu^{k}W_\nu^{i}$ are the Yang-Mills
field strengths associated to $\rm SU(2)_L$ gauge symmetry, and $g'$ is the $U(1)_{Y}$ coupling constant. Moreover, $\tilde{W}_{\mu\nu}^{j}=\frac{1}{2}\epsilon_{\mu\nu\rho\sigma}W^{j \rho\sigma}$ denotes the dual field strength tensor, while the Hermitian derivative operator $\overset{\text{\scriptsize$\leftrightarrow$}}{D}\hspace{0.0000000001cm}^\mu$ is defined as
$h\,\overset{\text{\scriptsize$\leftrightarrow$}}{D}\hspace{0.0000000001cm}^\mu f=h\,\overset{\text{\tiny$\rightarrow$}}{D}\hspace{0.0000000001cm}^\mu f-h\,\overset{\text{\tiny$\leftarrow$}}{D}\hspace{0.0000000001cm}^\mu f$. The coefficients $\bar c_i$ and $\tilde c_i$ appearing in Eq.~\eqref{LSILH} correspond to the Wilson coefficients of the SILH-basis effective operators. The terms that generate contributions to the $ZZh$ vertex are explicitly shown in Eq.~\eqref{LSILH}, whereas the ellipsis denotes additional terms which are not relevant for the present work.\\

Contributions to the $WWh$ vertex from the seesaw variant of radiatively-induced light-neutrino masses of Ref.~\cite{Pilaftsis} were recently investigated in Ref.~\cite{WWh}. In that work, $ZZh$-vertex ACs, in addition to those generated by the SILH basis, were identified and were found to be associated with effective-Lagrangian terms of mass dimension greater than 6. Such effectve-Lagrangian terms are
\begin{equation}
{\cal L}_{DW}=\frac{\overline{c}_{DW}}{m_W^4}
i\Big(
\big( D^\nu\Phi \big)^\dagger \, \sigma ^{j} \,D^\mu D^\rho\Phi\Big) \big(D_\mu W_{\nu\rho}\big)^{j}+{\rm H.c.},
\label{LDW}
\end{equation}
\begin{equation}
{\cal L}_{D \Phi}=\frac{g\hspace{0.1cm}\overline{c}_{D \Phi}}{2m_W^4}\Big(\Phi^\dagger D^\mu D^\rho\Phi\Big){\rm tr}\big\{ W^\nu\hspace{0.000001cm}_\rho W_{\nu\mu} \big\}+{\rm H.c.},
\label{L_Dphi}
\end{equation}
with the coefficients $\overline{c}_{D \Phi}$ and  $\overline{c}_{DW}$ dimensionless. The Lagrangian terms $\mathcal{L}_{D \Phi}$ and $\mathcal{L}_{DW}$ are invariant under $\rm SU(2)_L \otimes \rm U(1)_Y$ gauge group, and have canonical dimension 8. 
The effective-Lagrangian terms displayed in Eqs.~(\ref{LDW})-(\ref{L_Dphi}) are particularly interesting for the present work since, after electroweak symmetry breaking, they generate effective interactions involving a Higgs boson and two neutral gauge bosons, thereby inducing contributions to the $ZZh$ vertex. Thus, they provide additional sources of anomalous $ZZh$ couplings.\\

After implementation of the Brout-Englert-Higgs mechanism~\cite{Englert:1964et,Higgs:1964pj} to $\mathcal{L}_{\rm SILH}$, Eq.~(\ref{LSILH}), the resulting $ZZh$ couplings read
\begin{multline}
    \mathcal{L}_{\rm SILH}=\frac{g \,m_W\lambda_1}{2 c_W^{2}} Z_\mu Z^{\mu}h
-\frac{g\lambda_2}{4m_W}Z^{\mu\nu}Z_{\mu\nu} h\\-\frac{g\lambda_3}{m_W}\big( Z^{\nu}\partial^\mu Z_{\mu\nu}\big)h-\frac{g\tilde{\lambda}_1}{4m_W}Z_{\mu\nu}\tilde{Z}^{\mu\nu}h+\cdots,
\label{LSILH2}
\end{multline}
with $\lambda_1=1-\frac{\overline{c}_\Phi }{2}-2\overline{c}_T+8\overline{c}_{\gamma}\frac{s_W^4}{c_W^2}$, $\lambda_2=\frac{2}{c_W^2}\big(c_W^2\overline{c}_{\Phi W}+s_W^2\overline{c}_{\Phi B}-4s_W^4\overline{c}_{\gamma}\big)$, $\lambda_3=\frac{1}{c_W^2}\Big(\big(\overline{c}_{\Phi W}+\overline{c}_W\big)c_W^2+\big(\overline{c}_{\Phi B}+\overline{c}_B\big)s_W^2\Big)$, and $\tilde{\lambda}_1=\frac{2}{c_W^2}\big(c_W^2\tilde{c}_{\Phi W}+s_W^2\tilde{c}_{\Phi B}-4s_W^4\tilde{c}_{\gamma}\big)$. 
Also as a consequence of spontaneous breaking of electroweak symmetry, the mass-dimension-8 effective-Lagrangian terms $\mathcal{L}_{DW}$ and $\mathcal{L}_{D\Phi}$, respectively displayed in Eqs.~\eqref{LDW} and \eqref{L_Dphi}, are expressed as
\begin{eqnarray}
&&
{\cal L}_{DW}=
-\frac{g\lambda_4}{m_W^3}
Z^{\nu}\partial_\mu Z_{\nu\rho}\big(\partial^\mu\partial^\rho h\big)+\cdots,
\label{LDW2}
\end{eqnarray}
\begin{eqnarray}
&&
{\cal L}_{D \Phi}=-\frac{g\lambda_5}{m_W^3}Z^{\nu}\hspace{0.000001cm}_\rho Z_{\nu\mu}\,\big(\partial^\mu\partial^\rho h\big)+\cdots,
\label{LDphi2}
\end{eqnarray}
with $\lambda_4=\frac{\overline{c}_{DW}}{2g}$, and $\lambda_5=-\frac{c_W^2\overline{c}_{D\Phi}}{2g}$. Again, in the last two equations, the ellipses indicate the presence of further terms, corresponding to couplings other than $ZZh$.\\

\begin{figure}[h!]
  \includegraphics[width=.25\textwidth]{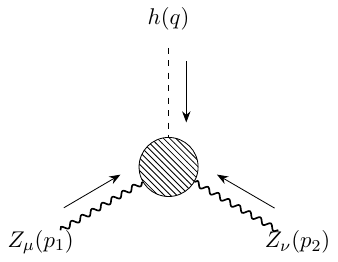}
  
                \caption{Conventions for the external momenta in the $ZZh$ vertex. All three external particles are taken off-shell.}
                \label{fig:hZZ}
\end{figure}

Now, let us consider an effective Lagrangian $\mathcal{L}_{ZZh}^{\rm eff}$, comprised by the terms explicitly displayed in Eqs.~(\ref{LSILH2})-(\ref{LDphi2}). This Lagrangian can be decomposed as $\mathcal{L}_{ZZh}^{\rm eff}=\mathcal{L}_{ZZh}^{\rm even}+\mathcal{L}_{ZZh}^{\rm odd}$, where $\mathcal{L}_{ZZh}^{\rm even}$ is $CP$-conserving, whereas $\mathcal{L}_{ZZh}^{\rm odd}$ accounts for $CP$-violating interactions. The corresponding terms, classified according to their $CP$-transformation properties, are given by
\begin{multline}
    \mathcal{L}_{ZZh}^{\rm even}=  \frac{g \,m_W}{2 c_W^{2}} \lambda_1Z_\mu Z^{\mu}h\\-\frac{g}{m_W}\Big(\frac{\lambda_2}{4}Z^{\mu\nu}Z_{\mu\nu}+\lambda_3Z^{\nu}\partial^\mu Z_{\mu\nu}\Big)h
  \\-\frac{g}{m_W^3}\Big(\lambda_4 Z^{\nu}\partial_\mu Z_{\nu\rho}+\lambda_5Z^{\nu}\hspace{0.000001cm}_\rho Z_{\nu\mu}\Big)\partial^{\mu}\partial^{\rho}h,
\end{multline}
\begin{equation}
    \mathcal{L}_{ZZh}^{\rm odd}=-\frac{g\tilde{\lambda}_1}{4m_W}Z_{\mu\nu}\tilde{Z}^{\mu\nu}h.
\end{equation}
Using the conventions of Fig.~\ref{fig:hZZ} we introduce the vertex function corresponding to the effective-Lagrangian $\mathcal{L}_{ZZh}^{\rm eff}$. Assuming all external particles to be off shell and imposing the 4-momentum conservation condition $p_1 + p_2 + q = 0$, the vertex-function $\Gamma^{\rm eff}_{\mu\nu}$ can be expressed in terms of the two independent 4-momenta $p_1$ and $p_2$. As we did with the effective Lagrangian $\mathcal{L}^\textrm{eff}_{ZZh}$, we split this vertex function into $CP$-even and $CP$-odd contributions, $\Gamma^{\rm eff}_{\mu\nu}= \Gamma^{\rm even}_{\mu\nu} + \Gamma^{\rm odd}_{\mu\nu}$, with the explicit expressions given by
    \begin{multline}
     \Gamma^{\rm even}_{\mu\nu}=g
\bigg(
\frac{m_W}{c_W^2}\lambda_1g_{\mu\nu}-\frac{\lambda_2}{m_W}\big( p_{1\nu}p_{2\mu}-g_{\mu\nu}\,p_1\cdot p_2 \big)\\-\frac{\lambda_3}{m_W}\big( p_{1\mu}p_{1\nu}+p_{2\mu}p_{2\nu}-g_{\mu\nu}p_1^2-g_{\mu\nu}p_2^2 \big)\\-\frac{\lambda_4}{m_W^3}
\Big( p_{1\mu}p_{1\nu}\big( p_1^2+p_1\cdot p_2 \big)+p_{2\mu}p_{2\nu}\big( p_2^2+p_1\cdot p_2 \big)\\
-g_{\mu\nu}\big( \big( p_2^2+p_1\cdot p_2 \big)^2+\big( p_1^2+p_1\cdot p_2 \big)^2 \big)
+p_{1\nu}p_{2\mu}(p_1+p_2)^2\Big)\\+\frac{2\lambda_5}{m_W^3}\Big( p_2^2\,p_{1\mu}p_{1\nu}
+p_1^2\,p_{2\mu}p_{2\nu}-(p_1\cdot p_2)p_{1\mu}p_{2\nu}-p_1^2\,p_2^2\,g_{\mu\nu}\\-\big(p_1^2+p_1\cdot p_2+p_2^2\big)\big(p_1\cdot p_2g_{\mu\nu}-p_{1\nu}p_{2\mu}\big) \Big)\bigg),
\label{Geven}
\end{multline}
\begin{equation}
      \Gamma_{\mu\nu}^{\rm odd}=
-\frac{g\tilde\lambda_1}{m_W}\epsilon_{\mu\nu\alpha\beta}\,p_1^\alpha p_2^\beta.
\label{Godd}
\end{equation}
\subsection{Virtual Majorana-neutrinos contributions to $ZZh$ at one-loop}
The main purpose of this subsection is to describe and discuss the analytical calculation of the one-loop contributions to the $ZZh$ vertex arising from the seesaw-mechanism variant presented in Ref.~\cite{Pilaftsis}. \\

Differences between Dirac and Majorana neutrinos manifest in the structure of fermion-field contractions and, as a direct consequence, in the topologies of the Feynman diagrams contributing to a given amplitude. In particular, the Majorana condition allows for additional fermion contractions, as dictated by Wick's theorem~\cite{Wick}, leading to a distinct set of Feynman rules~\cite{DMneutrinos,DMneutrinos2,Denner:1992me,Gates:1987ay,Haber:1984rc}.
Within the Majorana formalism, two different vertices must be taken into account for the $n_{j}n_{i}h$ couplings. For instance, the Lagrangian $\mathcal{L}_{hnn}$, which we display in Eq.~(\ref{Lhnn}), has the form $\mathcal{L}_{hnn}=-i\sum_{i}\sum_{j} h\, \overline{n}_{j} \Gamma_{ji} n_{i}$, from which the corresponding Feynman rules 
\begin{align}
     \begin{gathered}
            \includegraphics[scale=.4]{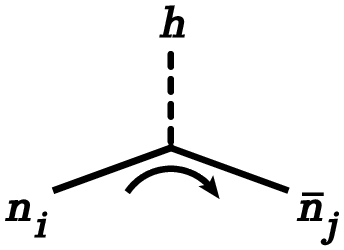}
    \end{gathered}&=\Gamma_{ji}, \label{Diracvertex} \\
     \begin{gathered}
            \includegraphics[scale=.4]{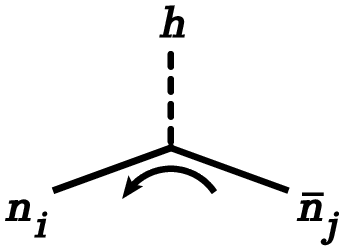}
    \end{gathered}&=C\, \Gamma_{ij}^\textrm{T} \,C^{-1},
\label{Majoranavertex}
\end{align}
are derived. In these equations, $\Gamma_{ij}^{T}$ denotes the transpose of $\Gamma_{ji}$ and $C$ is the charge-conjugation matrix. The Feynman rule in Eq.~(\ref{Diracvertex}) applies to both Dirac and Majorana neutrinos, while that of Eq.~(\ref{Majoranavertex}) is specific to the Majorana-neutrino case. Their diagrammatic difference is reflected in the orientation of the fermion-line arrows, which define a reference fermion flow~\cite{DENNER}. A similar situation occurs in the case of the  $\overline{n}_{j}n_{i}Z$ vertex, which also admits two different structures:
\begin{align}
   \begin{gathered}
            \includegraphics[scale=.4]{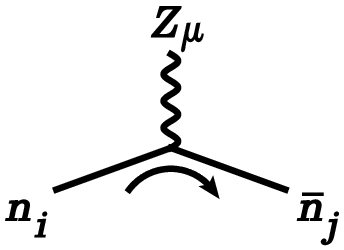}
    \end{gathered}&=\Gamma^{'}_{ji}, \label{Diracvertex2} \\
    \begin{gathered}
            \includegraphics[scale=.4]{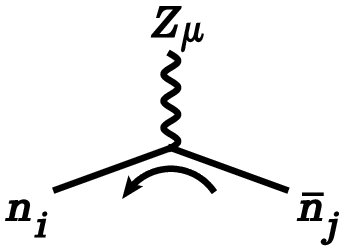}
    \end{gathered}&=C \,\Gamma^{'\textrm{T}}_{ij}\, C^{-1}, 
\label{Majoranavertex2}
\end{align}
where $\Gamma^{'}_{ji}$ is the Feynman rule associated to the couplings shown in Eq.~(\ref{LNC}).\\

The amplitude to calculate is given by
\begin{equation}
     \Gamma^{ZZh}_{\mu\nu}=\sum_{i=1}^{6}\sum_{j=1}^{6}\sum_{k=1}^{6}\Gamma^{ijk}_{\mu\nu},
\end{equation}
with the partial-amplitude contribution $\Gamma^{ijk}_{\mu\nu}$ diagrammatically expressed as
     \begin{eqnarray}
     &&
         \Gamma^{ijk}_{\mu\nu}=\begin{gathered}
            \includegraphics[scale=0.35]{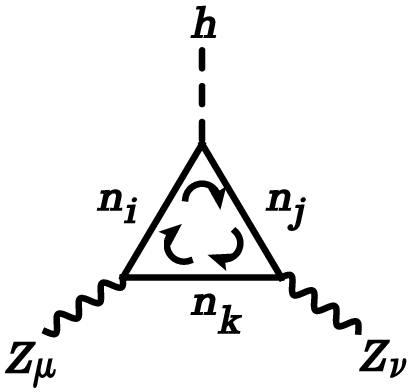}
    \end{gathered}+\begin{gathered}
            \includegraphics[scale=0.35]{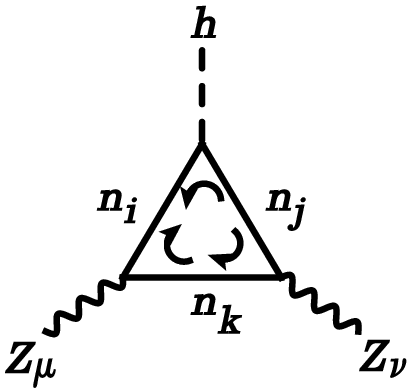}
    \end{gathered}
    \nonumber \\ \hspace{0.4cm} &&
    +\begin{gathered}
            \includegraphics[scale=0.35]{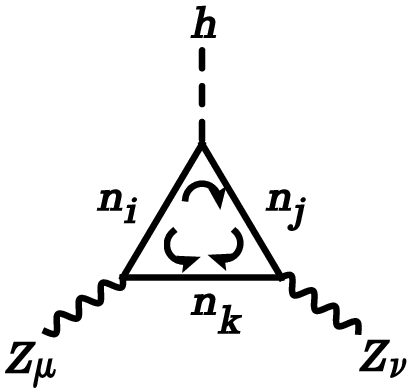}
    \end{gathered}+\begin{gathered}
            \includegraphics[scale=0.35]{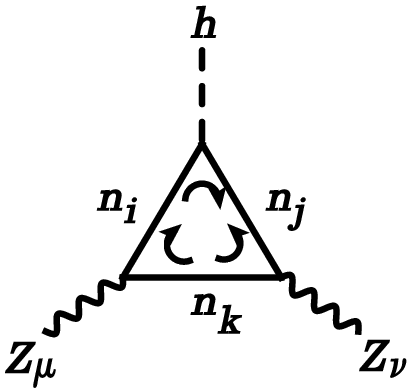}
    \end{gathered}+\text{BS}.
    \label{feynmandiagrams}
     \end{eqnarray}
The first diagram is of Dirac type, as it involves only Dirac-like $n_{j}n_{i}h$ vertex insertions, as defined in Eq.~(\ref{Diracvertex}), whereas the remaining three diagrams are of Majorana type, since they require the insertion of at least one Majorana $n_{j}n_{i}h$ or $n_{j}n_{i}Z$ coupling, given in Eqs.~(\ref{Majoranavertex})–(\ref{Majoranavertex2}). The acronym BS, standing for ``Bose symmetry'', indicates, in Eq.~\eqref{feynmandiagrams}, that further contributing diagrams obtained by the implementation of BS, through interchange of the two external $Z$ bosons, have been also considered.
Let us point out that, by virtue of the properties of the charge-conjugation matrix $C$, the analytical expressions arising from the Majorana-type diagrams are found to coincide with those of their Dirac-type counterparts.\\

Note that each of the one-loop diagrams contributing to the partial amplitude $\Gamma^{ijk}_{\mu\nu}$ displayed in Eq.~(\ref{feynmandiagrams}), including those generated by the BS prescription, exhibits a superficial degree of divergence equal to 1, thus signaling the potential appearance of ultraviolet divergences. With this in mind, we employ the method of dimensional regularization~\cite{Bollini,THOOFT}. Within this framework, spacetime dimensions are analytically continued to $D=4-\epsilon$, with $\epsilon\to 0$, which modifies the loop integrals as $\int \frac{d^4k}{(2\pi)^4}\to \mu_{R}^{4-D}\int \frac{d^Dk}{(2\pi)^D}$, where $\mu_{R}$ denotes the renormalization scale. The analytical evaluation of the corresponding Feynman diagrams is then carried out using the Passarino–Veltman tensor reduction method~\cite{PASSARINO1979151,DEVARAJ1998483}, implemented in \textsc{Mathematica}, by Wolfram, through the packages \textsc{FeynCalc}~\cite{FeynCalc1,FeynCalc2,FeynCalc3} and \textsc{Package-X}~\cite{Patel_2015}. \\

The analytical evaluation of the partial amplitudes $\Gamma_{\mu\nu}^{ijk}$, Eq.~\eqref{feynmandiagrams}, leads to a Lorentz structure that can be directly matched onto the effective $ZZh$ vertex parametrizations introduced in Eqs.~(\ref{Geven})--(\ref{Godd}), which were derived from the SILH Lagrangian supplemented with the relevant mass-dimension-8 operators. This matching procedure allows us to identify the one-loop Majorana-neutrino contributions to the ACs $\lambda_{i}$ and $\tilde{\lambda}_1$ by extracting the coefficients of the corresponding Lorentz structures. The resulting one-loop contributions to the ACs $\lambda_j$, with $j\neq 1$, and $\tilde{\lambda}_1$ are found to be
ultraviolet finite.
The only remaining ultraviolet divergence appears in $\lambda_1$, which is expected, since this coupling has a tree-level SM counterpart and can therefore be renormalized. The present investigation focuses on the one-loop contributions to ACs, not present in the SM at the tree level, so from here on we ignore the $\lambda_1$ coupling. In what follows, we denote the new-physics contributions to the ACs $\lambda_j$, with $j\neq 1$, and $\tilde{\lambda}_1$ by $L_\textrm{NP}^{(+)}$  and  $L_\textrm{NP}^{(-)}$. Specifically, $L_\textrm{NP}^{(+)}$ corresponds to the $CP$-even ACs $\lambda_j$, whereas $L_\textrm{NP}^{(-)}$ is associated with the $CP$-odd AC $\tilde{\lambda}_1$. All these quantities can be written in the generic form

\begin{widetext}
    \begin{multline}
    L_\textrm{NP}^{(\pm)}=\sum_{i=1}^{6}\sum_{j=1}^{6}\sum_{k=1}^{6}\bigg(L_{n_i\,n_j\,n_k}^{(1)}\Big(m_{n_i}\big(C_{n_i n_j}C_{n_j n_k}C_{n_k n_i}\pm C_{n_i n_j}^{*}C_{n_j n_k}^{*}C_{n_k n_i}^{*} \big)+m_{n_j}\big(C_{n_i n_j}^{*}C_{n_j n_k}C_{n_k n_i}\pm C_{n_i n_j}C_{n_j n_k}^{*}C_{n_k n_i}^{*} \big)\Big)\\
    +L_{n_i\,n_j\,n_k}^{(2)}\Big(m_{n_i}\big(C_{n_i n_j}C_{n_j n_k}C_{n_k n_i}^{*}\pm C_{n_i n_j}^{*}C_{n_j n_k}^{*}C_{n_k n_i} \big)+m_{n_j}\big(C_{n_i n_j}^{*}C_{n_j n_k}C_{n_k n_i}^{*}\pm C_{n_i n_j}C_{n_j n_k}^{*}C_{n_k n_i} \big)\Big)
    \bigg),
    \label{LNP}
\end{multline}
\end{widetext}
where $L^{(1)}_{n_i\,n_j\,n_k}$ and $L^{(2)}_{n_i\,n_j\,n_k}$ are functions of masses and squared off-shell external momenta, expressed in terms of Passarino–Veltman scalar 2-point and 3-point functions. While individual 2-point functions contain ultraviolet-divergent pieces, these cancel out in the combinations contributing to $L^{(\pm)}_\textrm{NP}$, consistently with the ultraviolet finiteness of the AC contributions discussed above, whereas the scalar 3-point functions entering the calculation are ultraviolet finite.

\section{Estimations and discussion of results}
\label{estimations}
This section aims at presenting a numerical evaluation of the contributions arising from Majorana neutrinos, as defined within the framework of Ref.~\cite{Pilaftsis}, to the form factors characterizing the $ZZh$ vertex. The analysis is based on the general off-shell $ZZh$ vertex result previously derived and focuses on specific on-shell configurations, each associated with a relevant physical process employed to estimate the magnitude of these contributions.\\

Neutrino oscillations provide direct evidence for non-zero masses of light neutrinos, in contrast with the assumptions of the SM. However, this phenomenon does not provide information on the absolute neutrino-mass scale. Cosmological observations have set a stringent upper bound on the sum of light-neutrino masses, $\sum_j m_{\nu_j}<0.12$ eV, at the 95\% confidence level~\cite{Alam,refId0}. In addition, the KATRIN Collaboration has reported an upper bound on the effective electron-antineutrino mass, which translates into the constraint $m_{\nu_j}\lesssim 0.45$ eV at the 90\% confidence level~\cite{Katrin}. This bound is independent of cosmological assumptions and applies regardless of whether neutrinos are Majorana or Dirac fermions. Although oscillation data require the three light-neutrino masses to be nondegenerate, the measured mass-squared differences, $\Delta m_{21}^{2} \sim 10^{-5}$ $\rm eV^{2}$~\cite{Abe_2024} and $|\Delta m_{32}^{2}|\sim 10^{-3}$ $\rm eV^{2}$~\cite{Minos,Nova,T2k,Icecube,Sk,km3net,Reno,DayaBay,Nova2,dayabay2}, are sufficiently small that treating the light-neutrino masses as degenerate does not affect the numerical results of the present calculation.\\

Note that in the radiative neutrino-mass generation mechanism of Ref.~\cite{Pilaftsis} light-neutrino masses vanish at tree level. In the analytical structure of Eq.~(\ref{LNP}), we implement this condition to the neutrino-mass factors explicitly shown, which originate from the tree-level Lagrangian (Higgs–neutrino interactions, described in Eq.~(\ref{Lhnn})) and which are therefore set to 0 whenever the mass corresponds to a light neutrino. On the other hand, the light-neutrino masses involved in internal propagators correspond to radiatively generated masses. In our numerical analysis, we adopt a degenerate light-neutrino mass spectrum and set $m_{\nu_1}=m_{\nu_2}=m_{\nu_3}\equiv m_{\nu}$, which we fix to $m_\nu=0.45$ eV, consistent with the current experimental upper bounds discussed above. Moreover, within the framework of the model considered for the present phenomenological investigation, the mass spectrum of the heavy states is constrained to be quasi-degenerate~\cite{Pilaftsis}, and we therefore take $m_{N_1}=m_{N_2}=m_{N_3}\equiv m_N$ throughout this work.\\

We begin our discussion by decomposing each sum over neutral leptons as $\sum_{j=1}^{6}=\sum_{\nu_j}+\sum_{N_j}$, where $\nu_j$ runs over the light-neutrino fields $\nu_1,\ \nu_2,\,\nu_3$, while $N_j$ runs over the three heavy-neutrino fields $N_1,\ N_2,\ N_3$. In this way, the triple sum in Eq.~(\ref{LNP}) can be rewritten as
\begin{multline}
    \sum_{i=1}^{6}\sum_{j=1}^{6}\sum_{k=1}^{6}=\sum_{\nu_i,\nu_j,\nu_k}+\sum_{\nu_i,\nu_j,N_k}+\sum_{\nu_i,N_j,\nu_k}+\sum_{\nu_i,N_j,N_k}\\
    +\sum_{N_i,\nu_j,\nu_k}+\sum_{N_i,\nu_j,N_k}+\sum_{N_i,N_j,\nu_k}+\sum_{N_i,N_j,N_k}.
\label{sumatoria}
\end{multline}
Building on the previous discussion regarding the distinction between the masses of light neutrinos at different levels, the ACs $L_{NP}^{(\pm)}$ receive no contributions from the first two terms in the decomposition shown in Eq.~(\ref{sumatoria}). On the other hand, the Hermiticity of the $6\times6$ matrix $\mathcal{C}$, defined in Eq.~(\ref{Cmatrix}), allows us to write the new-physics contributions, $L_\textrm{NP}^{(\pm)}$, as
\begin{multline}
    L_\textrm{NP}^{(+)}=2\,m_N\Big(
    L_{N\nu\nu}^{(1)}\,\text{Re}\,\text{Tr}\big\{\mathcal{C}_{\nu N}\mathcal{C}_{N\nu}\mathcal{C}_{\nu\nu}\big\}
    \\
    +\big(L_{\nu N\nu}^{(1)}+L_{\nu N\nu}^{(2)}+L_{N\nu\nu}^{(2)}\big)\,\text{Re}\,\text{Tr}\big\{\mathcal{C}_{N\nu}^{T}\mathcal{C}_{N\nu}\mathcal{C}_{\nu\nu}\big\}
    \\+\big(L_{\nu NN}^{(1)}+2L_{NN\nu}^{(2)}\big)\,\text{Re}\,\text{Tr}\big\{\mathcal{C}_{N\nu}\mathcal{C}_{N\nu}^{T}\mathcal{C}_{NN}\big\}\\
    +\big(L_{NN\nu}^{(1)}+L_{\nu NN}^{(2)}+L_{N\nu N}^{(2)}\big)\,\text{Re}\,\text{Tr}\big\{\mathcal{C}_{N\nu}\mathcal{C}_{\nu N}\mathcal{C}_{NN}^{T}\big\}\\
    +\big(L_{N\nu N}^{(1)}+L_{NN\nu}^{(1)}\big)\,\text{Re}\,\text{Tr}\big\{\mathcal{C}_{N\nu}\mathcal{C}_{\nu N}\mathcal{C}_{NN}\big\}\\
    +L_{NNN}^{(1)}\,\text{Re}\,\text{Tr}\big\{\mathcal{C}_{NN}\mathcal{C}_{NN}\big(\mathcal{C}_{NN}+\mathcal{C}_{NN}^{T}\big)\big\}\\
    +2L_{NNN}^{(2)}\,\text{Re}\,\text{Tr}\big\{\mathcal{C}_{NN}\mathcal{C}_{NN}\mathcal{C}_{NN}^{T}\big\}\Big),
\label{LNPCPeven}
\end{multline}
\begin{multline}
    L_\textrm{NP}^{(-)}
    \\
    =2im_N\Big(\big(L_{\nu N\nu}^{(1)}+L_{\nu N\nu}^{(2)}+L_{N\nu\nu}^{(2)}\big)\textrm{Im}\,\text{Tr}\big\{\mathcal{C}_{N\nu}\mathcal{C}_{\nu\nu}\mathcal{C}_{N\nu}^{T}\big\}\\+\big(L_{\nu NN}^{(1)}+2L_{\nu NN}^{(2)}\big)\textrm{Im}\,\text{Tr}\big\{\mathcal{C}_{N\nu}^{T}\mathcal{C}_{NN}\mathcal{C}_{N\nu}\big\}\Big).
\label{LNPCPodd}
\end{multline}
As it follows from Eq.~(\ref{LNPCPodd}), in contrast to the $CP$-even case, the $CP$-odd contribution $L_\textrm{NP}^{(-)}$ originates exclusively from interference terms involving both light- and heavy-neutrino sectors, and therefore does not admit contributions arising solely from heavy-neutrino exchange.\\

As shown in Eqs.~(\ref{LNPCPeven})–(\ref{LNPCPodd}), the new-physics contributions $L_\textrm{NP}^{(\pm)}$ are in part determined by the matrix $\mathcal{C}$, which can be written, following Ref.~\cite{Pilaftsis}, in terms of a complex $3\times 3$ matrix $\xi$. Since a fully general parametrization would involve a large number of free parameters, we adopt a simplified setup aimed at estimating the size of the effects under consideration. We therefore write $\xi=\hat{\rho}\,X$, where $\hat{\rho}$ is a real parameter setting the overall scale of $\xi$, and $X$ is a dimensionless complex matrix normalized such that its largest entry has unit modulus. In particular, we choose the texture $X=e^{i\phi}\cdot\mathbbm{1}_{3}$, so that 
\begin{equation}
   \xi=\hat{\rho}\,e^{i\phi}\cdot\mathbbm{1}_{3}. 
\label{texturaxi}
\end{equation}
Although this choice is evidently not general, we have checked that alternative textures do not appreciably affect our numerical results. Previous analyses~\cite{WWA,WWZ,ZZZ,WWh} have shown that the choices $\hat{\rho}=0.58$ and $\hat{\rho}=0.65$ lead to potentially observable effects. Accordingly, we fix $\hat{\rho}=0.65$, which is consistent with current CMS bounds implying $m_N \gtrsim 700~\text{GeV}$~\cite{PhysRevLett.120.221801}.\\

A global fit combining data from the Large Electron-Positron (LEP) Collider and the LHC was performed in Ref.~\cite{Ellis_2018}, in the framework of the Standard Model Effective Field Theory (SMEFT). That analysis constrains the $CP$-even Wilson coefficients of the SILH Lagrangian, yielding $\overline{c}_{\Phi W}=(0.2\,\pm\,1.4)\,\times\,10^{-2}$, $\overline{c}_{\Phi B}=(-1.3\,\pm\,1.8)\,\times\,10^{-2}$, $\overline{c}_{W}=(-1.9\,\pm\,2.4)\,\times\,10^{-2}$, $\overline{c}_{B}=(2.2\,\pm\,2.4)\,\times\,10^{-2}$ and $\overline{c}_{\gamma}=(-1\,\pm\,0.6)\,\times\,10^{-3}$ at 95\% CL. These results translate into the following bounds on the $CP$-even coefficients of our vertex-function parametrization: 
\begin{equation}
   |\lambda_2^{\rm global\,fit}|\leqslant 3.5\,\times\,10^{-2},
   \label{l2globalfit}
\end{equation}
\begin{equation}
   |\lambda_3^{\rm global\,fit}|\leqslant 3.6\,\times\,10^{-2}.
   \label{l3globalfit}
\end{equation}
Projected sensitivities at future lepton colliders have also been investigated. In particular, Ref.~\cite{Denizli_2018} analyzed Higgs production via $e^{+}e^{-}\to h\nu\overline{\nu}$ at the CLIC with a center-of-mass energy (CME) of 380 GeV, obtaining the following projected constraints:
\begin{equation}
   |\lambda_2^{\rm CLIC}|\leqslant 9.3\,\times\,10^{-3},
   \label{l2clic}
\end{equation}
\begin{equation}
   |\lambda_3^{\rm CLIC}|\leqslant 3\,\times\,10^{-2}.
   \label{l3clic}
\end{equation}
Similar studies have been performed for other in-plans lepton facilities. In the context of the ILC, the process $e^{+}e^{-}\to Z\gamma h$, at $\sqrt{s}=500$ GeV, was considered in Ref.~\cite{Alam_2017}, leading to projected sensitivities of comparable order:
\begin{equation}
   |\lambda_2^{\rm ILC}|\leqslant 4.3\,\times\,10^{-3},
   \label{l2ilc}
\end{equation}
\begin{equation}
   |\lambda_3^{\rm ILC}|\leqslant 9\,\times\,10^{-3}.
   \label{l3ilc}
\end{equation}
Likewise, Ref.~\cite{Spor_2025} studied a future Muon Collider (MC) operating at $\sqrt{s}=10\,\rm{TeV}$, probing anomalous Higgs-gauge interactions through  $\mu^{+}\mu^{-}\to Z\gamma\gamma$, and obtained a projected sensitivity of
\begin{equation}
   |\lambda_3^{\rm MC}|\leqslant 6\,\times\,10^{-2}.
   \label{l3mc}
\end{equation}
Further in-plans lepton colliders are the Circular Electron Positron Collider (CEPC) and the Future Circular Collider in its first stage, centered in electron-positron collisions and usually referred to as FCC-ee. About the perspectives for these future lepton colliders, let us comment on Ref.~\cite{DeBlas:2019qco}, whose authors performed a global fit from electroweak and Higgs-boson processes in the framework of the SM effective field theory, taking into account the not only the ILC and the CLIC, but also the CEPC and the FCC-ee. This paper reports the experimental reach projected for these machines, at the $95\%$ confidence level, on factors $\frac{c_j}{\Lambda^2}$, with $\Lambda$ the high-energy scale of the physical description behind the effective field theory and $c_j$ representing the different Wilson coefficients in a given basis of effective-Lagrangian operators. Their bounds translate into limits on the $CP$-even ACs $\lambda_2$ and $\lambda_3$, considered for the present work. According to this comprehensive work, the CEPC and the FCC-ee are expected to set bounds as stringent as $10^{-3}$ on these ACs. Constraints of the same order of magnitude are also claimed to be reachable by the CLIC and the ILC.
\\

In addition to constraints on $CP$-even effects, limits on $CP$-violating interactions have been reported by the ATLAS Collaboration, obtained from the process $p\,p\to h\to \gamma\,\gamma$~\cite{ATLAS-CONF-2019-029}. The resulting limits, at 95\% confidence level, on the SILH Wilson coefficients are $\tilde{c}_{\Phi W}=\tilde{c}_{\Phi B}=(-0.1\,\pm\,6.4)\,\times\,10^{-2}$ and $\tilde{c}_{\gamma}=(0.75\,\pm\,3.55)\,\times\,10^{-4}$, which imply
\begin{equation}
    |\tilde{\lambda}_{1}^{\rm ATLAS}|\leqslant 1.6\,\times\,10^{-1}.
    \label{l1atlas}
\end{equation}
Future-lepton-collider sensitivities to $CP$-odd effects have also been explored. Ref.~\cite{Karadeniz_2020} estimates that the CLIC operating at $\sqrt{s}=3$ TeV could probe $CP$-violating interactions down to
\begin{equation}
    |\tilde{\lambda}_{1}^{\rm CLIC}|\leqslant 3.2\,\times\,10^{-2}.
    \label{l1clic}
\end{equation}
An investigation of the process $e^+e^-\to Zh\to\ell^+\ell^- h$, performed in the framework of the CEPC, leaded the authors of Ref.~\cite{Sha:2022bkt} to conclude that this collider will be able to establish the contraint
\begin{equation}
|\tilde\lambda_1^\textrm{CEPC}|\leqslant 1.20\times10^{-2}.
\end{equation}
at the $95\%$ confidence level. Similarly, a prospective MC has been analyzed in Ref.~\cite{Gurkanli_2025} through the process $\mu^{+}\mu^{-}\to h\nu_{\ell}\overline{\nu}_{\ell}\to b\overline{b}\nu_{\ell}\overline{\nu}_{\ell}$, leading to an expected sensitivity at the level
\begin{equation}
|\tilde{\lambda}_{1}^{\rm MC}|\leqslant 1\,\times\,10^{-2}.
\label{l1mc}
\end{equation}
For our numerical study, we consider two kinematic configurations of the $ZZh$ vertex. The first one corresponds to Higgsstrahlung, with one on-shell $Z$ boson and one off-shell $Z^{*}$, while the Higgs boson is taken on shell. The second configuration describes VBF, where the Higgs boson is produced via two off-shell $Z^{*}$ bosons. These configurations are analyzed, below, in connection with the corresponding production mechanisms.
\hfill\break
\subsection{Scenario 1: $Z^{*}Zh$}

\begin{figure}[h!]
  \includegraphics[width=.2\textwidth]{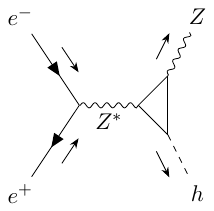}
  
                \caption{Generic one-loop diagram for light- and heavy-neutrino contributions to the $ZZh$ vertex in Higgsstrahlung, illustrated for $e^{-} e^{+} \to Zh$.}
                \label{fig:scenario1}
\end{figure}

The Higgsstrahlung process $e^{-}e^{+}\to Zh$, illustrated in Fig.~\ref{fig:scenario1} through the dominant $s$-channel exchange of an off-shell  $Z$ boson, constitutes one of the most relevant channels for precision measurements of the Higgs-boson mass, being highly sensitive to possible deviations from the SM while providing an exceptionally clean experimental environment. Higher-order corrections to Higgsstrahlung within the SM, including one- and multi-loop contributions, have been computed in detail for future linear colliders in several works~\cite{sym13071256,Jegerlehner_2005,Bondarenko,BELANGER2003252,Freitas,Liu_2014}.\\

Our estimations of the new-physics contributions $L_\textrm{NP}^{(\pm)}$ are performed over a CME range extending from the kinematic threshold for $Zh$ production up to 1 TeV, namely $m_{h}+m_{Z}\leqslant\sqrt{s}\leqslant 1$ TeV. In the Higgsstrahlung configuration, the Lorentz structure of the effective vertex function $\Gamma_{\mu\nu}^{\rm eff}$, after imposing kinematic and transversality conditions, prevents an independent study of the $CP$-even coefficients. Instead, only specific linear combinations of the relevant ACs can be probed. Accordingly, we define 
\begin{equation}
\Delta_{2}=\lambda_2+\frac{(p_{1}+p_{2})^{2}}{m_{W}^{2}}\lambda_{4}-\frac{2}{m_{W}^{2}}\Big(p_{1}^{2}+p_{1}\cdot p_{2}+p_{2}^{2}\Big)\lambda_{5},
\end{equation}
\begin{equation}
\Delta_{3}=\lambda_3+\frac{p_{2}^{2}+p_{1}\cdot p_{2}}{m_{W}^{2}}\lambda_{4}-\frac{2p_{1}^{2}}{m_{W}^{2}}\lambda_{5}.
\end{equation}
These combinations encode contributions from operators of different canonical dimensions. While $\lambda_{2}$ and $\lambda_{3}$ arise from mass dimension-6 operators, the coefficients $\lambda_{4}$ and $\lambda_{5}$ originate from mass dimension-8 terms. 
Let us remark that the AC new-physics contributions $L_\textrm{NP}^{(\pm)}$ are, in general, complex quantities. However, for our discussion we rather focus on their absolute values, namely $|\Delta_2|$ and $|\Delta_3|$. 
\begin{figure}[h!]
\center
\includegraphics[width=7.5cm]{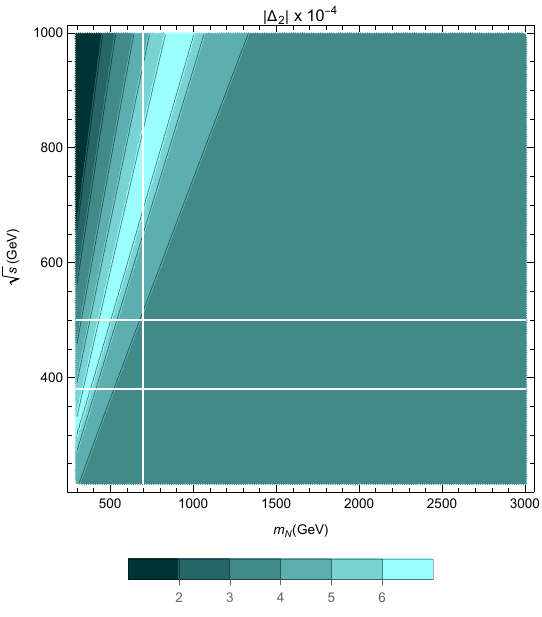}
\vspace{0.3cm}
\\
\includegraphics[width=7.5cm]{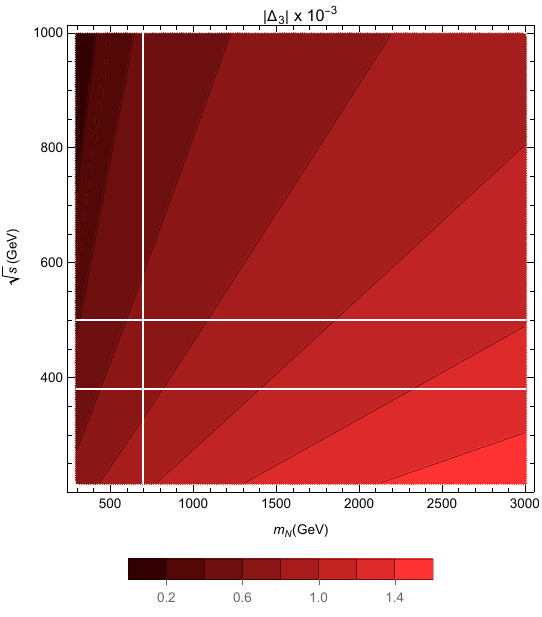}
\caption{\label{Delta23Sc1} Contributions to the $CP$-even combinations $|\Delta_2|$ (upper panel) and $|\Delta_3|$ (lower panel) in the Higgsstrahlung configuration, shown in the $(m_N,\sqrt{s})$ plane for fixed $\hat{\rho}=0.65$. The parameter ranges are $300\,{\rm GeV}\leqslant m_N\leqslant 3\,{\rm TeV}$ and $m_h+m_Z\leqslant\sqrt{s}\leqslant 1\,{\rm TeV}$. Contributions as large as $|\Delta_2|\sim 10^{-4}$ and $|\Delta_3|\sim 10^{-3}$ are obtained.}
\end{figure}
\\

Fig.~\ref{Delta23Sc1} displays the corresponding contributions in the parameter space $(\,m_N,\,\sqrt{s}\,)$, with $300\,{\rm GeV} \leqslant m_N\leqslant3\,{\rm TeV}$ and $m_h+m_Z\leqslant\sqrt{s}\leqslant1\,{\rm TeV}$. The color gradients indicate the magnitude of the contributions, with lighter regions corresponding to larger values. The color bars are scaled to orders of $10^{-4}$ and $10^{-3}$ for $|\Delta_2|$ and $|\Delta_3|$, respectively.  Consequently, the combinations $\Delta_i$ depend on the heavy-neutrino mass and $\sqrt{s}$, $\Delta_i=\Delta_i(m_N,\sqrt{s})$. Horizontal lines at $\sqrt{s}=380\,{\rm GeV}$ and $\sqrt{s}=500\,{\rm GeV}$ denote representative benchmark values of the CME for future lepton colliders, such as the CLIC and the ILC, while the vertical line at $m_N=700\,{\rm GeV}$ corresponds to the minimum heavy-neutrino mass allowed by our choice $\hat{\rho}=0.65$, consistent with the CMS analysis of heavy neutral leptons~\cite{PhysRevLett.120.221801}. Inspection of the upper panel of Fig.~\ref{Delta23Sc1} shows that $\big| \Delta_2 \big|$ might reach values as large as $~\sim10^{-4}$. This graph also suggests that points in which $\sqrt{s}\sim m_N$ correspond to maxima of $\big| \Delta_2 \big|$. This is in fact the case, which can be seen by considering CME values near $\sqrt{s}=m_N$ and then expanding $\Delta_2$ as $\Delta_2\simeq \Delta_2^{(0)}+\Delta_2^{(1)}\big( \sqrt{s}-m_N \big)+\Delta_2^{(2)}\big( \sqrt{s}-m_N \big)^2$, where $\Delta_2^{(k)}=\frac{1}{k!}\frac{d^k\Delta_2}{d\sqrt{s}\hspace{0.0001cm}^k}\Big|_{\sqrt{s}=m_N}$. Then, from this expansion, the critical points $\sqrt{s}_\textrm{max.}=m_N-\frac{\Delta_2^{(1)}}{2\Delta_2^{(2)}}$ are determined and the ratio $\frac{\Delta_2^{(1)}}{2\Delta_2^{(2)}}$ is verified to be small enough in  order for $\sqrt{s}_\textrm{max.}\approx m_N$ to approximately correspond to the maxima of $\Delta_2$. Since, as commented earlier, the masses of heavy neutral leptons are constrained to be $700\,\textrm{GeV}\lesssim m_N$, the occurrence of $\sqrt{s}_\textrm{max.}\approx m_N$ means that the sub-region in the upper graph of Fig.~\ref{Delta23Sc1} in which $700\,\textrm{GeV}\lesssim\sqrt{s}$ has its relevance, as it includes all the maxima of the $|\lambda_2|$ AC contribution. Regarding the $|\Delta_3|$ contribution, the lower panel of Fig.~\ref{Delta23Sc1} shows that it varies between approximately $10^{-4}$ and $10^{-3}$.\\

For $CP$-violating effects, the numerical analysis was carried out over the same parameter space. Our estimates for $|\tilde{\lambda}_1|$ are of order $10^{-15}$. Comparing this result with the bounds given in Eqs.~(\ref{l1atlas})–(\ref{l1mc}), we find that, even in the most optimistic scenario, our prediction is approximately 13 orders of magnitude smaller than the projected sensitivities of future lepton colliders. The strong suppression of $|\tilde{\lambda}_1|$ may be related to the absence of a contribution exclusively associated with heavy-neutrino exchange, as indicated by Eq.~(\ref{LNPCPodd}), in contrast with the $CP$-conserving case where heavy-neutrino effects provide the dominant contribution.
Additionally, we tested several matrix textures for $\mathcal{C}$ and observed no significant variation in the order of magnitude of $|\tilde{\lambda}_1|$.

\subsection{Scenario 2: $Z^{*}Z^{*}h$}
\begin{figure}[h!]
  \includegraphics[width=.3\textwidth]{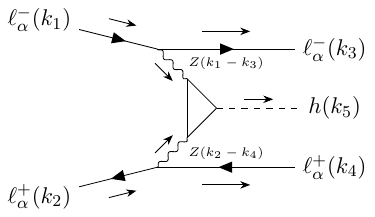}
  
                \caption{Generic one-loop diagram for light- and heavy-neutrino contributions to the $ZZh$ vertex in neutral-current VBF, illustrated for $\ell^{-}\ell^{+}\to \ell^{-}\ell^{+}h$.}
                \label{fig:scenario2}
\end{figure}

We now turn to Higgs production via neutral-current VBF, 
$\ell^{-}\ell^{+}\to Z^{*}Z^{*}\to \ell^{-}\ell^{+}h$, 
at lepton colliders. In this configuration, the Higgs boson is produced through the fusion of two off-shell $Z$ bosons, allowing for a direct probe of the $ZZh$ vertex structure. The corresponding one-loop diagram for light- and heavy-neutrino corrections to the $ZZh$ vertex is shown in Fig.~\ref{fig:scenario2}.
The kinematic configuration of neutral-current VBF does not impose the same restrictions as in the Higgstrahlung case on the Lorentz structure of $\Gamma_{\mu\nu}^{\rm eff}$, thereby allowing for an independent analysis of each of the $CP$-even ACs. Furthermore, the presence of 2 virtual gauge bosons prevents the CME of the collision from directly fixing the allowed values of the off-shell  $Z$-boson squared 4-momenta, $(k_1-k_3)^{2}$ and $(k_2-k_4)^{2}$. The kinematic limits on these momentum transfers can be derived following the same phase-space decomposition procedure employed in Ref~\cite{WWh}. In particular, by factorizing the three-body phase space into a $2\to2$ subprocess followed by a 2-body decay and imposing energy-momentum conservation, one finds that the allowed range for the squared 4-momenta of the virtual $Z$ bosons is given by
\begin{equation}
 -s\Big(1-\frac{m_{h}^{2}}{s}\Big)\leqslant k_{13}^{2}\leqslant0,
 \label{k13}
\end{equation}
\begin{equation}
 -s\Big(1-\frac{m_{h}^{2}}{s}\Big)\leqslant k_{24}^{2}\leqslant0,
 \label{k24}
\end{equation}
where $s$ denotes the squared CME of the initial lepton pair, and we have defined $k_{13}^{2}=(k_1-k_3)^{2}$ and $k_{24}^{2}=(k_2-k_4)^{2}$.\\

\begin{figure*}[t]
\centering

\begin{subfigure}[t]{0.45\textwidth}
\centering
\includegraphics[width=\textwidth]{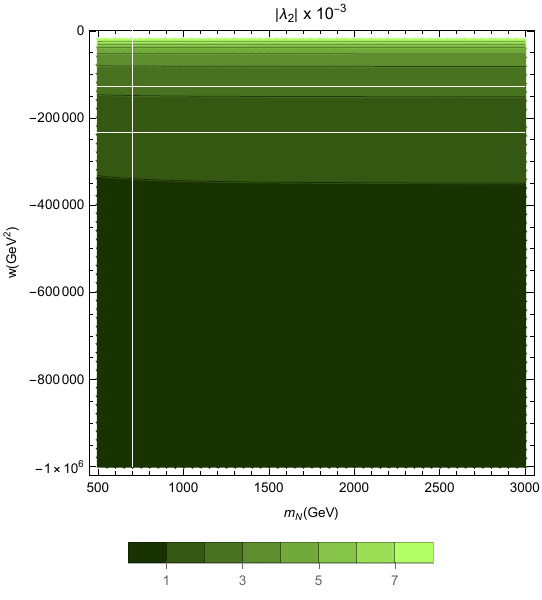}
\end{subfigure}
\hfill
\begin{subfigure}[t]{0.45\textwidth}
\centering
\includegraphics[width=\textwidth]{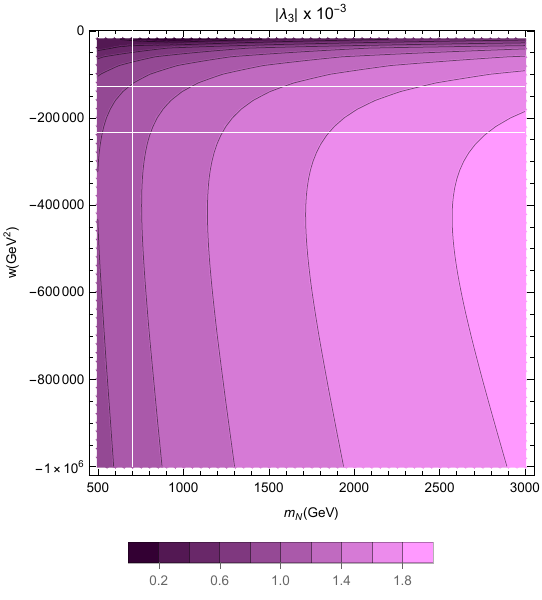}
\end{subfigure}

\vspace{0.5cm}

\begin{subfigure}[t]{0.45\textwidth}
\centering
\includegraphics[width=\textwidth]{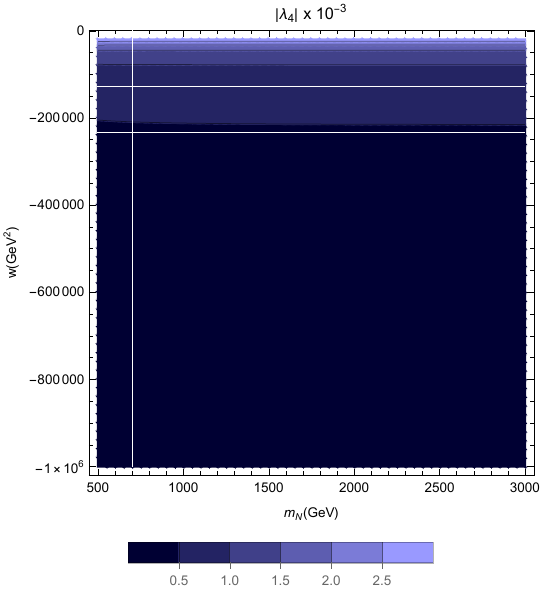}
\end{subfigure}
\hfill
\begin{subfigure}[t]{0.45\textwidth}
\centering
\includegraphics[width=\textwidth]{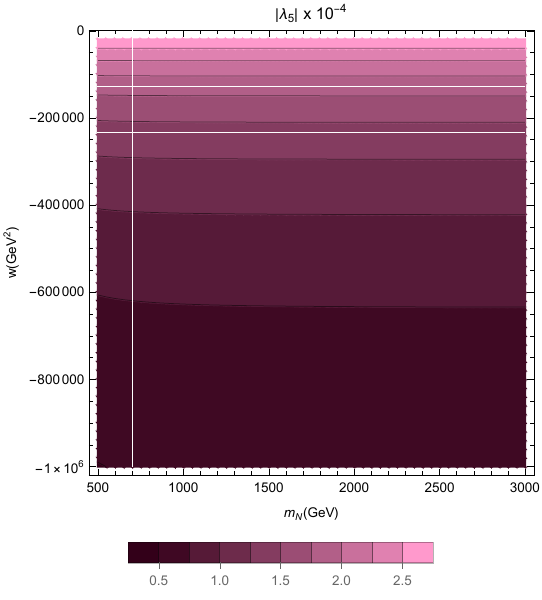}
\end{subfigure}

\caption{Contour plots of the $CP$-even form factors in neutral-current VBF, displayed in the $(m_N,\,w)$ plane for the benchmark choice $\delta_1=\frac{1}{5}$ and $\delta_2=\frac{1}{30}$, and fixed $\hat{\rho}=0.65$. The upper-left (upper-right) panel shows $|\lambda_2|$ ($|\lambda_3|$), while the lower-left (lower-right) panel displays $|\lambda_4|$ ($|\lambda_5|$). The ranges $500\,{\rm GeV} \leq m_N \leq 3\,{\rm TeV}$ and $-1\,{\rm TeV}^2 \leq w \leq 0\,{\rm TeV^{2}}$ have been considered. }
\label{fig:CPeven_ZZ_VBF}
\end{figure*}

We begin our discussion of numerical estimations by considering the $CP$-preserving effects. According to Eqs.~(\ref{k13})–(\ref{k24}), the squared 4-momenta $k_{13}^{2}$ and $k_{24}^{2}$ are independent kinematic variables. In what follows, we parametrize the virtual $Z$-boson squared 4-momenta as $k_{13}^{2}=\delta_1 w$ and $k_{24}^{2}=\delta_2 w$. Here,
\begin{equation}
-s\left(1-\frac{m_{h}^{2}}{s}\right)\leqslant w \leqslant 0,
\end{equation}
\begin{equation}
0\leqslant \delta_1 \leqslant 1,
\end{equation}
\begin{equation}
0\leqslant \delta_2 \leqslant 1.
\end{equation}
A numerical exploration of the $(\delta_1,\,\delta_2)$ parameter space indicates that the $CP$-even form factors attain their largest values for configurations in which the two virtual gauge bosons have significantly different momentum transfers. In particular, the benchmark choice $\delta_1=\frac{1}{5}$ and $\delta_2=\frac{1}{30}$ corresponds to a representative point within this region of enhanced contributions. Under this choice, we estimate the contributions $|\lambda_2|$, $|\lambda_3|$, $|\lambda_4|$, and $|\lambda_5|$, which are displayed in Fig.~\ref{fig:CPeven_ZZ_VBF}. 
The plots are presented in the $(m_N,\,w)$ parameter space, with $500\,{\rm GeV} \leqslant m_N\leqslant 3\,{\rm TeV}$ and $-1\,{\rm TeV}^2 \leqslant w\leqslant 0\,{\rm TeV}^{2}$. As before, the colors characterizing each region correspond to different values of the moduli $|\lambda_j|$ for $j=2,3,4,5$, with lighter shades indicating larger contributions. The vertical lines at $m_N=700\,{\rm GeV}$, included in the plots, represent the minimal allowed heavy-neutrino mass according to Ref.~\cite{PhysRevLett.120.221801}. Additionally, the horizontal lines correspond to the minimal values of $w$ for lepton colliders with CME $\sqrt{s}=380\,{\rm GeV}$ and $\sqrt{s}=500\,{\rm GeV}$, as determined from Eqs.~(\ref{k13})–(\ref{k24}).

\begin{figure}[t]
\centering
\includegraphics[width=3.5cm]{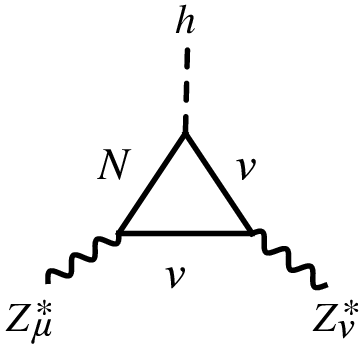}
\hspace{1cm}
\includegraphics[width=3.5cm]{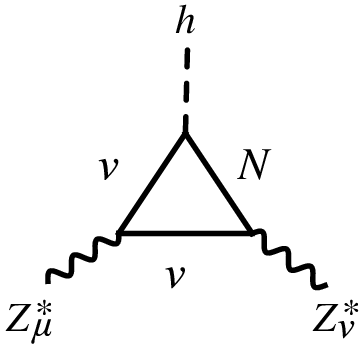}
\\
\textcolor{Red}{($N\nu\nu$)}\hspace{3.8cm}\textcolor{Red}{($\nu N\nu$)}
\vspace{0.7cm}
\\
\includegraphics[width=3.5cm]{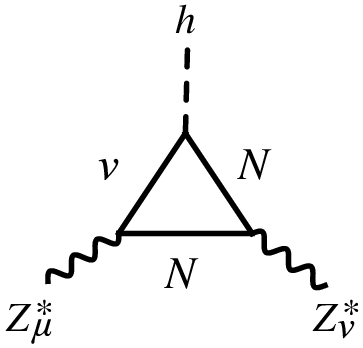}
\hspace{1cm}
\includegraphics[width=3.5cm]{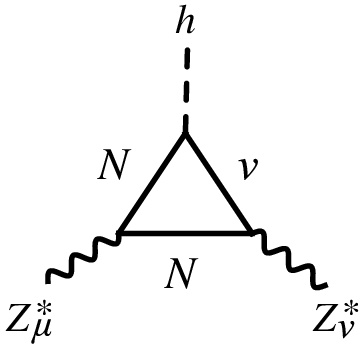}
\\
\textcolor{Red}{($\nu NN$)}\hspace{3.8cm}\textcolor{Red}{($N\nu N$)}
\vspace{0.7cm}
\\
\includegraphics[width=3.5cm]{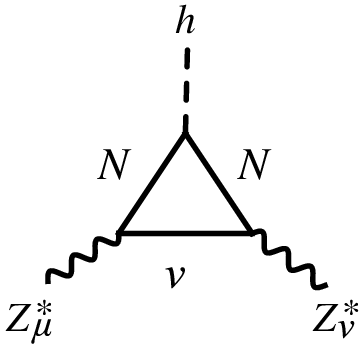}
\hspace{1cm}
\includegraphics[width=3.5cm]{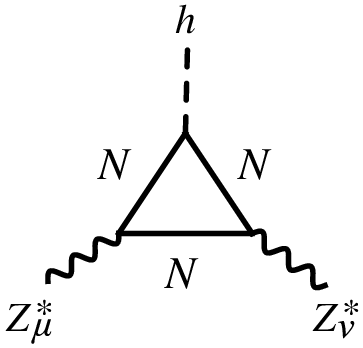}
\\
\textcolor{Red}{($NN\nu$)}\hspace{3.8cm}\textcolor{Red}{($NNN$)}
\caption{Classification of contributing diagrams, according to their loop-fermion lines: light neutrinos are represented by $\nu$, whereas $N$ stands for heavy neutral leptons. A label for each type of diagram has been included.}
\label{VBFdiags}
\end{figure}

\begin{figure*}[t]
\centering

\begin{subfigure}[t]{0.45\textwidth}
\centering
\includegraphics[width=\textwidth]{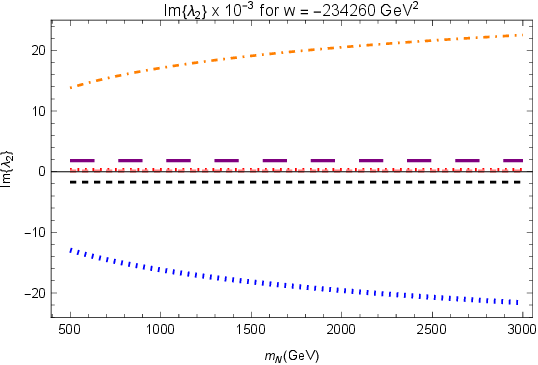}
\end{subfigure}
\hfill
\begin{subfigure}[t]{0.45\textwidth}
\centering
\includegraphics[width=\textwidth]{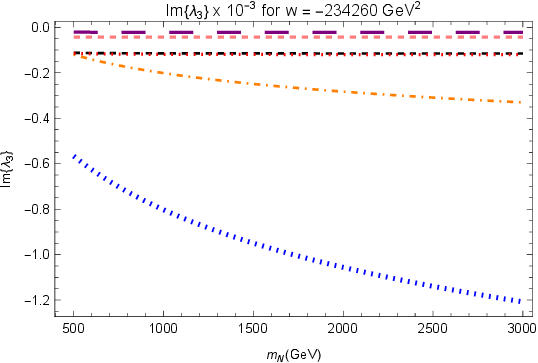}
\end{subfigure}

\vspace{0.5cm}

\begin{subfigure}[t]{0.45\textwidth}
\centering
\includegraphics[width=\textwidth]{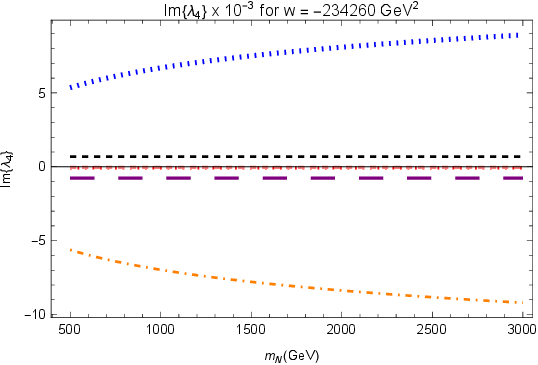}
\end{subfigure}
\hfill
\begin{subfigure}[t]{0.45\textwidth}
\centering
\includegraphics[width=\textwidth]{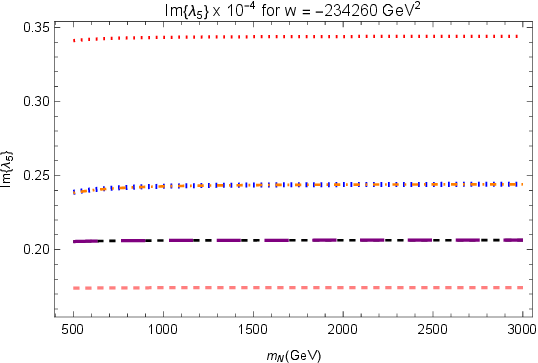}
\end{subfigure}
\vspace{0.4cm}

\centering
\small

\begin{tabular}{ccc}

\begin{tikzpicture}[baseline=-0.6ex]
\draw[blue,dotted,very thick] (0,0)--(0.7,0);
\end{tikzpicture}
$\operatorname{Im}\{\lambda_{j}^{\nu N\nu}\}$

&
\begin{tikzpicture}[baseline=-0.6ex]
\draw[black,dashed,thick] (0,0)--(0.7,0);
\end{tikzpicture}
$\operatorname{Im}\{\lambda_{j}^{\nu NN}\}$

&
\begin{tikzpicture}[baseline=-0.6ex]
\draw[orange,dash dot,thick] (0,0)--(0.7,0);
\end{tikzpicture}
$\operatorname{Im}\{\lambda_{j}^{N\nu\nu}\}$

\\[2mm]

\begin{tikzpicture}[baseline=-0.6ex]
\draw[violet,dash pattern=on 6pt off 4pt,very thick]
(0,0)--(0.7,0);
\end{tikzpicture}
$\operatorname{Im}\{\lambda_{j}^{N\nu N}\}$

&
\begin{tikzpicture}[baseline=-0.6ex]
\draw[red,dotted,very thick]
(0,0)--(0.7,0);
\end{tikzpicture}
$\operatorname{Im}\{\lambda_{j}^{NN\nu}\}$

&
\begin{tikzpicture}[baseline=-0.6ex]
\draw[pink,dashed,very thick]
(0,0)--(0.7,0);
\end{tikzpicture}
$\operatorname{Im}\{\lambda_{j}^{NNN}\}$

\end{tabular}

\vspace{0.3cm}
\caption{Virtual-neutrino contributions to the imaginary parts of $CP$-even AC $\lambda_j$. Each graph shows, separately,  the contributions from different types of diagrams, in accordance with Fig.~\ref{VBFdiags}. Labels for the different contributions appear below the graphs. The choice of parameters $\delta_1=\frac{1}{5}$, $\delta_2=\frac{1}{30}$, $w=-234260\,\textrm{GeV}^2$ has been used to plot the graphs, which has been done for heavy-neutrino-mass values within the range $0.5\,\textrm{TeV}\leqslant m_N\leqslant3\,\textrm{TeV}$.}
\label{Imaginaryparts}
\end{figure*}

We observe that the contributions to the $CP$-even ACs $|\lambda_2|$ and $|\lambda_3|$ can reach values as large as $\sim 10^{-3}$. These results suggest that the corresponding new-physics effects could lie close to the expected sensitivity of the ILC. In contrast, the contributions to the form factors $|\lambda_4|$ and $|\lambda_5|$ are typically of order $10^{-4}$. Numerically, the resulting contributions are of the same order of magnitude as those obtained in the Higgsstrahlung channel, highlighting the complementary role of neutral-current VBF in probing the effective $ZZh$ interaction. As an additional remark, we have tried using other asymmetry choices of $\delta_1$ and $\delta_2$, obtaining $CP$-even AC contributions of the same order of magnitude as those obtained for the benchmark configuration. An exception occurs in the symmetric limit $\delta_1=\delta_2$, an aspect we find worth of further comment. Let us first note that the graphs of Fig~\ref{fig:CPeven_ZZ_VBF} suggest that the behavior of the $|\lambda_3|$ contribution, with respect to the heavy-neutrino mass $m_N$, is notably different from all other $CP$-even AC contributions. In the VBF Higgs-production process $\ell^-\ell^+\to Z^*Z^*\to\ell^-\ell^+ h$, which defines the framework behind these graphs, both $Z$-boson external lines are off the mass shell, from which relations among contributing diagrams emerge. To better grasp this observation, we refer the reader to Fig~\ref{VBFdiags},
which shows the type of one-loop diagrams contributing to $ZZh$, in the context of Higgs production by VBF. 
In this figure, the diagrams have been classified according to which neutral leptons, either light or heavy neutrinos, appear in the loop lines. Each label ``$\nu$'' or ``$N$'' then represents, in a generic manner (no need for neutrino indices), a light or a heavy neutrino, respectively. Note that each of the diagrams in Fig.~\ref{VBFdiags} is accompanied by a label, colored in red. Now note that diagrams $(N\nu\nu)$ and $(\nu N\nu)$ can be obtained from each other by interchanging the loop lines attached to the external Higgs field. And the same relation applies for diagrams $(\nu NN)$ and $(N\nu N)$. On the other hand, such a relation does not hold in the case of diagrams $(NN\nu)$ and $(NNN)$. Therefore, as long as the the squared momenta of the off-shell $Z$ bosons are close to each other, the $ZZh$ AC contributions from two diagrams which are similar in the sense of this relation are expected to be close to each other, possibly differing by an overall sign. To illustrate this, let us first recall that the AC contributions are complex valued, so $\lambda_j=\textrm{Re}\{\lambda_j\}+i\,\textrm{Im}\{\lambda_j\}$. Bear in mind that for the parameter ranges of interest the imaginary part of any $\lambda_j$ is dominant over its real part. Then consider Fig.~\ref{Imaginaryparts},
which comprises four graphs, each displaying a set of curves corresponding to contributions, from each type of diagram alone, to the imaginary part of an AC $\lambda_j$. For the execution of these graphs, we have fixed $w=-234260\,\textrm{GeV}^2$, whereas the values $\delta_1=\frac{1}{5}$ and $\delta_2=\frac{1}{30}$, previously used to plot the graphs of Fig.~\ref{fig:CPeven_ZZ_VBF}, have been kept. In the case of the graphs representing $\textrm{Im}\{ \lambda_2 \}$ (upper-left panel) and $\textrm{Im}\{ \lambda_4 \}$ (lower-left panel), the afore-discussed relation among types of diagrams, given by the interchange of neutral lepton lines in the $hnn$ vertex, yields a destructive sum from related diagrams, thus producing suppressed total contributions to $\textrm{Im}\{ \lambda_2 \}$ and $\textrm{Im}\{ \lambda_4 \}$. In fact, such a cancellation yields an exact null contribution from the sum of related diagrams if $\delta_1=\delta_2$ is taken, as in that case the momenta of the external off-shell $Z$ bosons exactly coincide and thus an overall sign is the only difference among contributions. Moreover, in this case the total contributions $\textrm{Im}\{ \lambda_2 \}$ and $\textrm{Im}\{ \lambda_4 \}$ are completely determined by the subdominant diagrams $(NN\nu)$ and $(NNN)$, which vary slowly with respect to $m_N$. In the case of $\textrm{Im}\{ \lambda_3 \}$ and $\textrm{Im}\{ \lambda_5 \}$, the graphs placed in the upper-right and lower-right panels of Fig.~\ref{Imaginaryparts} show that all the contributions produced by the diagrams are negative, so no suppression driven by a destructive contribution seems to occur, in contrast with the other $CP$-even AC contributions $\lambda_2$ and $\lambda_4$. 
Even though the contributions to both $\textrm{Im}\{ \lambda_3 \}$ and $\textrm{Im}\{ \lambda_5 \}$ from all the diagrams are negative, the behavior of these AC contributions differs near $\delta_1=\delta_2$, namely,
the sums of pairs of related diagrams to $\textrm{Im}\{ \lambda_3 \}$ completely vanish, whereas the analogous contributions to $\textrm{Im}\{ \lambda_5 \}$ remain nonzero. While the $\textrm{Im}\{ \lambda_3 \}$ graph (upper-right panel of Fig.~\ref{Imaginaryparts}) suggests the absence of a destructive sum of contributions among diagrams, it turns out that as $\delta_1$ and $\delta_2$ approach to each other, eventually one of the related diagrams switches its overall sign, then giving rise to the elimination of the total contribution at $\delta_1=\delta_2$. This produces to a suppression of the total contribution, which is now determined by the subdominant, non-related, diagrams $(NN\nu)$ and $(NNN)$. As commented before, this is not the case of $\lambda_5$. Another feature of the $CP$-even AC contributions to be emphasized is that the relation among diagrams under the interchange of $hnn$ fermion lines is completely absent if the process of Higgs production by Higgstahlung, discussed thoughout the previous subsection, is considered. This makes sense, since in that case the squared momentum of the off-shell $Z$ boson varies, whereas the squared momentum of the on shell $Z$ boson is fixed to $m_Z^2$ by the on-shell condition. This is the reason why the comportment of $|\Delta_2|$ with respect to $m_N$, seemingly stable for a large region in the upper panel of Fig.~\ref{Delta23Sc1}, cannot be the result of a destructive contribution among diagrams. And neither can $|\Delta_3|$ experience an alike suppression.
\\

We now turn to the $CP$-violating effects. After implementing a degenerate heavy-neutrino mass spectrum, together with the texture of the matrix $\xi$ given in Eq.~(\ref{texturaxi}), the $CP$-odd contribution can be expressed as
\begin{equation}
\tilde{\lambda}_1 = \Omega_{\nu N}\,\sin(2\phi),
\end{equation}
where $\Omega_{\nu N}$ is a function of neutral-lepton masses. The explicit dependence on the phase $\phi$ shows that the form factor $\tilde{\lambda}_1$ reaches its maximal value for $\phi=\frac{\pi}{4}$. Moreover, in contrast to the $CP$-preserving contributions, the degree of asymmetry between the momentum transfers $k_{13}^{2}$ and $k_{24}^{2}$ does not play a significant role in determining the order of magnitude of $|\tilde{\lambda}_1|$, since the $CP$-odd contribution is already strongly suppressed over the explored parameter space. In the $(m_N,\,w)$ parameter space, our estimates for $|\tilde{\lambda}_1|$ lie in the range $10^{-16}$–$10^{-15}$, which is consistent with the results obtained in the Higgsstrahlung scenario.

\section{Summary and conclusions}
\label{summary}
In this work, we have calculated, estimated, and discussed the one-loop contributions from both light and heavy Majorana neutrinos to the $ZZh$ vertex within a variant of the type-I seesaw mechanism in which the light-neutrino masses vanish at tree level and are generated radiatively. \\

The resulting structure of the $ZZh$ vertex function contains contributions compatible with the Lorentz-covariant structures generated by effective operators of canonical dimension 6, together with additional terms associated with higher-dimensional interactions. Besides the SM tree-level coupling, the vertex is characterized by $CP$-conserving and $CP$-violating ACs. While the $CP$-even contributions are described by the coefficients $\lambda_2$ and $\lambda_3$, first emerging from mass-dimension-6 effective-Lagangian terms, and by $\lambda_4$ and $\lambda_5$, associated to nonrenormalizable terms of canonical dimension 8, the CP-odd structure is encoded in $\tilde{\lambda}_1$. All AC contributions are found to be ultraviolet finite.\\

To estimate the magnitude of the AC contributions, we considered two scenarios. The first corresponds to Higgsstrahlung, in which one $Z$ boson is off shell, while the other $Z$ boson and the Higgs boson are on shell. This scenario was analyzed in the context of future lepton colliders. The kinematic configuration in this case does not allow for an independent study of the $CP$-even ACs contributions; however, specific linear combinations of the relevant couplings can be probed. In this setup, we find $CP$-conserving contributions as large as $\sim 10^{-3}$.
By contrast, the $CP$-odd contribution is strongly suppressed, being of order $10^{-15}$, well below the projected future experimental sensitivities.\\

The second scenario corresponds to Higgs production via neutral-current VBF, $\ell^{-}\,\ell^{+}\to\ell^{-}\,\ell^{+} h$, where the kinematics allows for an independent study of the $CP$-even coefficients. Among the free parameters involved in the AC contributions are the squared 4-momenta of the virtual $Z$ bosons, which we varied in accordance with constraints imposed by the phase-space of the corresponding cross section. We explored the case in which the squared 4-momenta of the virtual $Z$ bosons are different and found $CP$-even contributions as large as $\sim 10^{-3}$ for $|\lambda_2|$ and $|\lambda_3|$, potentially within the projected sensitivity of future lepton colliders. In addition, we obtain contributions of order $10^{-4}$ for $|\lambda_4|$ and $|\lambda_5|$.

\section*{Acknowledgements}
\noindent
The authors acknowledge financial support from SECIHTI (México). H.V. acknowledges support from the program ``\textit{Becas Nacionales para Estudios de Posgrado 2024-2}'' (support number 4037802). M.S. acknowledges support from the SECIHTI program "\textit{Estancias Posdoctorales por México}".

\appendix*

\section{The model}
This Appendix aims at providing a brief description of the model given in Ref.~\cite{Pilaftsis}, considered for the present phenomenological investigation. We star by considering the inclusion of 3 right handed singlet neutrino fields, $\nu'_{1,R}, \,\nu'_{2,R},\,\nu'_{3,R}$, in which case the following renormalizable terms are allowed by SM gauge symmetry:
\begin{equation}
\mathcal{L}^\nu_\textrm{mass}=-\frac{1}{2}\overline{{\nu_R'}^\textrm{c}}m_\textrm{M}\nu'_R-\overline{L_L}\tilde{\phi}\mathcal{Y}^\nu\nu'_R+\textrm{H.c.},
\label{Lnumass}
\end{equation}
where $m_\textrm{M}$ is a symmetric complex $3\times3$ matrix, assumed to originate from a stage of spontaneous symmetry breaking at some high energy scale $\Lambda$, $\mathcal{Y^\nu}$ is a $3\times3$ Yukawa matrix, and $\tilde{\phi}=i\sigma^2\phi^*$, with $\phi$ the SM Higgs doublet and $\sigma^2$ the imaginary Pauli matrix. Moreover, 
\begin{equation}
L_L=
\left(
\begin{array}{c}
L_{e,L}
\\
L_{\mu,L}
\\
L_{\tau,L}
\end{array}
\right),
\hspace{0.5cm}
\nu'_R=
\left(
\begin{array}{c}
\nu'_{1,R}
\\
\nu'_{2,R}
\\
\nu'_{3,R}
\end{array}
\right),
\end{equation}
where $L_{\alpha,L}$, with $\alpha=e,\mu,\tau$, denotes the SM left-handed $\textrm{SU}(2)_L$ lepton doublets. Also note that ${\nu'_R}^\textrm{c}$ is the charge-conjugated field associated to $\nu'_R$. After electroweak symmetry breaking of the electroweak gauge group at $v=246\,\textrm{GeV}$, Eq.~\eqref{Lnumass} can be rearranged into the expression
\begin{equation}
\mathcal{L}^\nu_\textrm{mass}=-\frac{1}{2}
\left(
\begin{array}{cc}
\overline{\nu'_L} & \overline{{\nu'_R}^\textrm{c}}
\end{array}
\right)
\left(
\begin{array}{cc}
0 & m_\textrm{D}
\vspace{0.2cm}
\\
m_\textrm{D}^\textrm{T} & m_\textrm{M}
\end{array}
\right)
\left(
\begin{array}{c}
{\nu'}^\textrm{c}_L
\\
\nu'_R
\end{array}
\right)+\textrm{H.c.}
\end{equation}
Implementation of a Takagi diagonalization~\cite{Teiji:TAKAGI192483}, driven by the unitary $6\times6$ matrix $\mathcal{U}_\nu$, conveniently written in terms of $3\times3$ block matrices $\mathcal{U}_{jk}$ as
\begin{equation}
\mathcal{U}=
\left(
\begin{array}{cc}
\mathcal{U}_{11} & \mathcal{U}_{12}
\vspace{0.2cm}
\\
\mathcal{U}_{21} & \mathcal{U}_{22}
\end{array}
\right)
\label{Ublocks}
\end{equation}
yields
\begin{equation}
\mathcal{U}^\textrm{T}
\left(
\begin{array}{cc}
0 & m_\textrm{D}
\vspace{0.2cm}
\\
m_\textrm{D}^\textrm{T} & m_\textrm{M}
\end{array}
\right)\mathcal{U}
=
\left(
\begin{array}{cc}
m_\nu & 0
\vspace{0.2cm}
\\
0 & m_N
\end{array}
\right)
\equiv M_n.
\label{takagidiagonalization}
\end{equation}
This diagonalization procedure comes along with the change of basis that defines the set of neutrino-mass eigenspinors, comprised by three light neutrino fields, $\nu_1,\nu_2,\nu_3$, and by three heavy-neutral-lepton fields, $N_1,N_2,N_3$. Regarding Eq.~\eqref{takagidiagonalization}, notice that the $3\times3$ diagonal matrix $m_\nu$ is the one characterizing the masses of light neutrinos $\nu_j$, whereas $m_N$, also $3\times3$ sized and diagonal, comprises the masses of the heavy neutral leptons $N_j$. As discussed in Ref.~\cite{Pilaftsis}, the assumption that the condition
\begin{equation}
\left(
\begin{array}{cc}
0 & m_\textrm{D}
\vspace{0.2cm}
\\
m^\textrm{T}_\textrm{D} & m_\textrm{M}
\end{array}
\right)
\left(
\begin{array}{c}
\mathcal{U}_{11}
\vspace{0.2cm}
\\
\mathcal{U}_{21}
\end{array}
\right)=0
\end{equation}
holds leads to the elimination of the tree-level light-neutrino masses, namely, $m_\nu=0$, whereas masses of heavy neutral leptons $N_j$ remains untouched and given as $m_N\sim\Lambda$.
\\

The inclusion of the right-handed neutrino singlets $\nu'_{j,R}$, together with the corresponding mass terms given in Eq.~\eqref{Lnumass}, lead to couplings of neutral leptons $\nu_j$ and $N_j$ with the SM particle content:
\begin{eqnarray}
&&
\mathcal{L}_{hnn}=\frac{g}{4m_W}h\,\overline{n}\Big( P_R(M_n\mathcal{C}^*+\mathcal{C}M_n)
\nonumber \\ && \hspace{1cm}
+P_L(M_n\mathcal{C}+\mathcal{C}^*M_n) \Big)n,
\label{LhnnApdx}
\end{eqnarray}
\begin{equation}
\mathcal{L}_{Znn}=\frac{g}{4c_W}Z_\mu\overline{n}\gamma^\mu\big( \mathcal{C}P_L-\mathcal{C}^*P_R \big)n,
\label{LZnnApdx}
\end{equation}
\begin{equation}
\mathcal{L}_{W\ell n}=\frac{g}{\sqrt{2}}W^-_\rho\overline{\ell}\,\mathcal{B}\,\gamma^\rho P_L n+\textrm{H.c.},
\label{Lwln}
\end{equation}
\begin{equation}
\mathcal{L}_{G_{W}\ell n}=\frac{g}{\sqrt{2}\,m_W}G_W^-\overline{\ell}\big(M_\alpha\,\mathcal{B}\,P_L - P_R\,\mathcal{B}\,M_n) n+\textrm{H.c.},
\label{LGwln}
\end{equation}
\begin{eqnarray}
&&
\mathcal{L}_{G_{Z} nn}=\frac{i\,g}{4m_W}G_{Z}\,\overline{n}\Big( P_R(M_n\mathcal{C}^*+\mathcal{C}M_n)
\nonumber \\ && \hspace{1cm}
-P_L(M_n\mathcal{C}+\mathcal{C}^*M_n) \Big)n,
\label{LGznnApdx}
\end{eqnarray}
where $h$ represents the Higgs-boson field and $Z_\mu$ stands for the $Z$-boson field. Moreover, $G_W^+$, $G_W^-$, and $G_Z$ stand for the pseudo-Goldstone bosons that emerge after electroweak symmetry breaking. Note that the Lagrangians $\mathcal{L}_{hnn}$ and $\mathcal{L}_{Znn}$ have been previously presented in the main text (see Eqs.~\eqref{Lhnn} and \eqref{LNC}). As we explained then, we have defined the $6\times1$ matrix
\begin{equation}
n=
\left(
\begin{array}{c}
\nu
\vspace{0.1cm}
\\
N
\end{array}
\right),
\end{equation}
where $\nu$ and $N$ are $3\times1$ matrices with entries $\nu_1=n_1, \nu_2=n_2, \nu_3=n_3$ and $N_1=n_4,N_2=n_5,N_3=n_6$, respectively. Also,
\begin{equation}
\ell=
\left(
\begin{array}{c}
\ell_{e,L}
\vspace{0.1cm}
\\
\ell_{\mu,L}
\vspace{0.1cm}
\\
\ell_{\tau,L}
\end{array}
\right).
\end{equation}
Regarding the tree-level elimination of light-neutrino masses, note that $m_\nu$, which is part of $M_n$, must be set to 0 in Eq.~\eqref{LhnnApdx}, that is, couplings of the Higgs boson and the light neutrinos are absent at the tree level, while they can be induced radiatively. Lepton charged currents, given by Eq.~\eqref{Lwln}, are characterized by the $3\times6$ matrix $\mathcal{B}=\big( \mathcal{B}_\nu,\mathcal{B}_N \big)$, constituted by the $3\times3$ matrices $\mathcal{B}_\nu$ and $\mathcal{B}_N$, defined, in terms of the $3\times3$ block matrices $\mathcal{U}_{jk}$ of Eq.~\eqref{Ublocks}, as
\begin{equation}
\mathcal{B}_\nu=\big(U_\textrm{PMNS}\,\mathcal{U}_{11}\big)^*,
\hspace{0.5cm}
\mathcal{B}_N=\big( U_\textrm{PMNS}\,\mathcal{U}_{12} \big)^*.
\end{equation}
The $\mathcal{B}$ matrix fulfills $\mathcal{B}\mathcal{B}^\dag={\bf 1}_3$, with ${\bf 1}_3$ the $3\times3$ identity matrix, and $\mathcal{B}^\dag\mathcal{B}=\mathcal{C}$, where the $6\times6$ matrix $\mathcal{C}$ defines the $hnn$ and $Znn$ couplings. Written in terms of $3\times3$ blocks, $\mathcal{C}$ reads
\begin{equation}
\mathcal{C}=
\left(
\begin{array}{cc}
\mathcal{C}_{\nu\nu} & \mathcal{C}_{\nu N}
\vspace{0.2cm}
\\
\mathcal{C}_{N\nu} & \mathcal{C}_{NN}
\end{array}
\right)
=
\left(
\begin{array}{cc}
\mathcal{U}^\textrm{T}_{11}\,\mathcal{U}^*_{11} & 
\mathcal{U}_{11}^\textrm{T}\,\mathcal{U}^*_{12}
\vspace{0.2cm}
\\
\mathcal{U}^\textrm{T}_{12}\,\mathcal{U}^*_{11}
&
\mathcal{U}^\textrm{T}_{12}\,\mathcal{U}^*_{12}
\end{array}
\right).
\end{equation}
\\

Neutrino masses are generated at the one-loop level by considering the following nonzero contributions to the 2-point function:
\begin{eqnarray}
&&
-i\Sigma_{\nu_j\nu_k}\big( \slashed{p} \big)=\sum_{k=1}^3
\Bigg(
\hspace{0.2cm}\begin{gathered}
\vspace{-0.1cm}
\includegraphics[width=2.6cm]{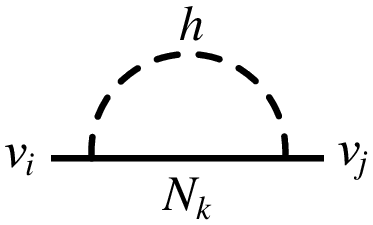}
\end{gathered}
\nonumber \\ && \hspace{1.5cm}
+
\begin{gathered}
\vspace{-0.1cm}
\includegraphics[width=2.6cm]{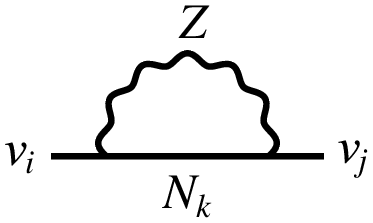}
\end{gathered}
+
\begin{gathered}
\vspace{-0.1cm}
\includegraphics[width=2.6cm]{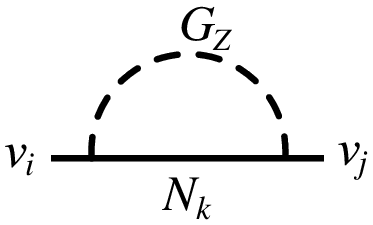}
\end{gathered}
\hspace{0.2cm}
\Bigg).
\nonumber \\ 
\end{eqnarray}
From this expression, the neutrino-mass contribution
\begin{eqnarray}
&&
\big( m_\nu^\textrm{1 loop} \big)_{jk}=\frac{g^2}{64\pi^2m_W^2}\sum_{i=1}^3\mathcal{C}_{\nu_jN_i}\mathcal{C}_{\nu_kN_i}m_{N_i}^3
\nonumber \\ &&
\times
\Big(
\frac{m_h^2}{m_{N_i}^2-m_h^2}\log\bigg( \frac{m_{N_i}^2}{m_h^2} \bigg)+\frac{3m_Z^2}{m_{N_i}^2-m_Z^2}\log\Big( \frac{m_{N_i}^2}{m_Z^2} \Big)
\Big)
\nonumber \\
\end{eqnarray}
is found~\cite{Pilaftsis}. We whish to emphasize two remarkable features of this result: (1) the contribution is ultraviolet finite and independent of the renormalization scale introduced as part of the dimensional-regularization method; (2) a careful calculation performed in the $R_\xi$ gauge shows that the contribution is independent of the gauge choice, in which case the mass is genuinely physical. Both qualities follow from the relation
\begin{equation}
\sum_{k=1}^3\mathcal{C}_{\nu_j N_k}\mathcal{C}_{\nu_i N_k}m_{N_k}=0
\label{lighnumasscondition}
\end{equation}
found to hold as long as light neutrino masses vanish at the tree level. Moreover, from the very same Eq.~\eqref{lighnumasscondition} the conclusion is reached that the masses of light neutrinos are tiny if the mass spectrum of heavy neutral leptons is quasi-degenerate.
\\

The unitary diagonalization matrix $\mathcal{U}$ can be expressed as~\cite{Pilaftsis}
\begin{equation}
\mathcal{U}=
\left(
\begin{array}{cc}
\big( \textbf{1}_3+\xi^*\xi^\textrm{T} \big)^{-\frac{1}{2}}
&
\xi^*\big( \textbf{1}_3+\xi^\textrm{T}\xi^* \big)^{-\frac{1}{2}}
\vspace{0.2cm}
\\
-\xi^\textrm{T}\big( \textbf{1}_3+\xi^*\xi^\textrm{T} \big)^{-\frac{1}{2}}
&
\big( \textbf{1}_3+\xi^\textrm{T}\xi^* \big)^{-\frac{1}{2}}
\end{array}
\right),
\end{equation}
where $\xi$ is a complex $3\times3$ matrix. If $\frac{v}{\Lambda}$ is assumed to be small, then $\xi$ can be approximated as $\xi \simeq m_\textrm{D}m_\textrm{M}^{-1}$. The introduction of the $\xi$ matrix allows one to express the matrices $\mathcal{B}$ and $\mathcal{C}$ as
\begin{equation}
\mathcal{B}_\nu=U_\textrm{PMNS}^*\big( \textbf{1}_3+\xi\xi^\dag \big)^{-\frac{1}{2}},
\end{equation}
\begin{equation}
\mathcal{B}_N=U_\textrm{PMNS}^*\,\xi\big( \textbf{1}_3+\xi^\dag\xi \big)^{-\frac{1}{2}},
\end{equation}
\begin{equation}
\mathcal{C}_{\nu\nu}=\big( \textbf{1}_3+\xi\xi^\dag \big)^{-1},
\end{equation}
\begin{equation}
\mathcal{C}_{\nu N}=\xi\big( \textbf{1}_3+\xi^\dag\xi \big)^{-1},
\end{equation}
\begin{equation}
\mathcal{C}_{N\nu}=\xi^\dag\big( \textbf{1}_3+\xi\xi^\dag \big)^{-1},
\end{equation}
\begin{equation}
\mathcal{C}_{NN}=\xi^\dag\xi\big( \textbf{1}_3+\xi^\dag\xi \big)^{-1}.
\end{equation}
Our numerical estimations are carried out with the assumption that the texture of the complex matrix $\xi$ is
\begin{equation}
\xi=e^{i\phi}\hat\rho\cdot\textbf{1}_3
\end{equation}
where $\hat\rho$ is a real number, such that $0<\hat\rho<1$, and $\textbf{1}_3$ is the $3\times3$ identity matrix. Under such circumstances, the matrices $\mathcal{B}$ and $\mathcal{C}$, characterizing the couplings of neutrinos with SM fields (see Eqs.~\eqref{LhnnApdx}-\eqref{LGznnApdx}), are expressed as
\begin{equation}
\mathcal{B}_{\nu}=\frac{1}{\sqrt{{1+\hat{\rho}^2}}}U_\textrm{PMNS}^*,
\end{equation}
\begin{equation}
\mathcal{B}_{\nu}=\frac{\hat{\rho}\,e^{i\phi}}{\sqrt{{1+\hat{\rho}^2}}}U_\textrm{PMNS}^*,
\end{equation}
\begin{equation}
    \mathcal{C}_{\nu\nu}=\frac{1}{1+\hat{\rho}²}\cdot\textbf{1}_3,
\end{equation}
\begin{equation}
    \mathcal{C}_{\nu N}=\frac{\hat{\rho}\,e^{i\phi}}{1+\hat{\rho}²}\cdot\textbf{1}_3,
\end{equation}
\begin{equation}
    \mathcal{C}_{N\nu }=\frac{\hat{\rho}\,e^{-i\phi}}{1+\hat{\rho}²}\cdot\textbf{1}_3,
\end{equation}
\begin{equation}
    \mathcal{C}_{NN}=\frac{\hat{\rho}^2}{1+\hat{\rho}²}\cdot\textbf{1}_3,
\end{equation}

\bibliographystyle{apsrev4-2}
\bibliography{MNZZh}

@article{Djouadi_2008,
   title={The anatomy of electroweak symmetry breaking},
   volume={457},
   ISSN={0370-1573},
   url={http://dx.doi.org/10.1016/j.physrep.2007.10.004},
   DOI={10.1016/j.physrep.2007.10.004},
   number={1–4},
   journal={Physics Reports},
   publisher={Elsevier BV},
   author={Djouadi, Abdelhak},
   year={2008},
   month=feb, pages={1–216} }

@article{Aad_2012,
   title={Observation of a new particle in the search for the Standard Model Higgs boson with the ATLAS detector at the LHC},
   volume={716},
   ISSN={0370-2693},
   url={http://dx.doi.org/10.1016/j.physletb.2012.08.020},
   DOI={10.1016/j.physletb.2012.08.020},
   number={1},
   journal={Physics Letters B},
   publisher={Elsevier BV},
   author={Aad, G. and \it{et al.} \rm(ATLAS Collaboration)},
   year={2012},
   month=sep, pages={1–29} }

@article{Chatrchyan_2012,
   title={Observation of a new boson at a mass of 125 GeV with the CMS experiment at the LHC},
   volume={716},
   ISSN={0370-2693},
   url={http://dx.doi.org/10.1016/j.physletb.2012.08.021},
   DOI={10.1016/j.physletb.2012.08.021},
   number={1},
   journal={Physics Letters B},
   publisher={Elsevier BV},
   author={Chatrchyan, S. and \it{et al.} \rm{(CMS Collaboration)}},
   year={2012},
   month=sep, pages={30–61} }

@article{Chatrchyan_2014,
   title={Measurement of the properties of a Higgs boson in the four-lepton final state},
   volume={89},
   ISSN={1550-2368},
   url={http://dx.doi.org/10.1103/PhysRevD.89.092007},
   DOI={10.1103/physrevd.89.092007},
   number={9},
   journal={Physical Review D},
   publisher={American Physical Society (APS)},
   author={Chatrchyan, S. and {\it et. al.} {\rm (CMS Collaboration)}},
   year={2014},
   month=may }

@article{Aad_2014,
   title={Measurement of the Higgs boson mass from the<mml:math xmlns:mml="http://www.w3.org/1998/Math/MathML" display="inline"><mml:mi>H</mml:mi><mml:mo stretchy="false">→</mml:mo><mml:mi>γ</mml:mi><mml:mi>γ</mml:mi></mml:math>and<mml:math xmlns:mml="http://www.w3.org/1998/Math/MathML" display="inline"><mml:mrow><mml:mi>H</mml:mi><mml:mo stretchy="false">→</mml:mo><mml:mi>Z</mml:mi><mml:msup><mml:mrow><mml:mi>Z</mml:mi></mml:mrow><mml:mrow><mml:mo>*</mml:mo></mml:mrow></mml:msup><mml:mo stretchy="false">→</mml:mo><mml:mn>4</mml:mn><mml:mo>ℓ</mml:mo></mml:mrow></mml:math>channels in<mml:math xmlns:mml="http://www.w3.org/1998/Math/MathML" display="inline"><mml:mi>p</mml:mi><mml:mi>p</mml:mi></mml:math>collisions at center-of-mass energies of 7 and 8 TeV with the ATLAS detector},
   volume={90},
   ISSN={1550-2368},
   url={http://dx.doi.org/10.1103/PhysRevD.90.052004},
   DOI={10.1103/physrevd.90.052004},
   number={5},
   journal={Physical Review D},
   publisher={American Physical Society (APS)},
   author={Aad, G. and {\it et. al} {\rm (ATLAS Collaboration)}},
   year={2014},
   month=sep }

@article{2016,
   title={Measurements of the Higgs boson production and decay rates and constraints on its couplings from a combined ATLAS and CMS analysis of the LHC pp collision data at s = 7 $$ \sqrt{s}=7 $$ and 8 TeV},
   volume={2016},
   ISSN={1029-8479},
   url={http://dx.doi.org/10.1007/JHEP08(2016)045},
   DOI={10.1007/jhep08(2016)045},
   number={8},
   journal={Journal of High Energy Physics},
   publisher={Springer Science and Business Media LLC},
   author={Aad, G. and {\it et. al} {\rm (ATLAS and CMS Collaboration)}},
   year={2016},
   month=aug }

@article{Aad:2915296,
      author        = "Aad, Georges and {\it et. al} {\rm (ATLAS Collaboration)}",
       title        = "{Measurements of $WH$ and $ZH$ production with Higgs boson
                       decays into bottom quarks and direct constraints on the
                       charm Yukawa coupling in $13\,\mathrm{TeV}$ $pp$ collisions
                       with the ATLAS detector}",
      archivePrefix = "arXiv",
      eprint        = "2410.19611",
      reportNumber  = "CERN-EP-2024-237",
      journal       = "JHEP",
      volume        = "2504",
      pages         = "075",
      year          = "2025",
      url           = "https://cds.cern.ch/record/2915296",
      doi           = "10.1007/JHEP04(2025)075",
}

@article{ATLAS-CONF-2017-055,
      author = "M. Aaboud and {\it et. al} {\rm (ATLAS Collaboration)}",
      journal ="J. High Energ. Phys.",
      title         = "{Search for heavy resonances decaying to a $W$ or $Z$ boson and a Higgs boson in final states with leptons and $b$-jets in 36.1~fb$^{-1}$ of $pp$ collision data at $\sqrt{s} = 13$~\TeV\ with the ATLAS detector}",
      institution   = "CERN",
      reportNumber  = "ATLAS-CONF-2017-055",
      address       = "Geneva",
      year          = "2017"
}

@article{2021,
   title={Measurements of WH and ZH production in the $$H \rightarrow b\bar{b}$$ decay channel in pp collisions at $$13\,\text {Te}\text {V}$$ with the ATLAS detector},
   volume={81},
   ISSN={1434-6052},
   url={http://dx.doi.org/10.1140/epjc/s10052-020-08677-2},
   DOI={10.1140/epjc/s10052-020-08677-2},
   number={2},
   journal={The European Physical Journal C},
   publisher={Springer Science and Business Media LLC},
   author={Aad, G. and {\it et. al} {\rm (ATLAS Collaboration)}},
   year={2021},
   month=feb }

@article{CMS2022,
   title={Measurement of the Higgs boson width and evidence of its off-shell contributions to ZZ production},
   volume={18},
   ISSN={1745-2481},
   url={http://dx.doi.org/10.1038/s41567-022-01682-0},
   DOI={10.1038/s41567-022-01682-0},
   number={11},
   journal={Nature Physics},
   publisher={Springer Science and Business Media LLC},
   author={Tumasyan, A. and {\it et. al} {\rm (CMS Collaboration)}},
   year={2022},
   month=oct, pages={1329–1334} }

@article{ATLAS2015,
   title={Constraints on the off-shell Higgs boson signal strength in the high-mass ZZ and WW final states with the ATLAS detector},
   volume={75},
   ISSN={1434-6052},
   url={http://dx.doi.org/10.1140/epjc/s10052-015-3542-2},
   DOI={10.1140/epjc/s10052-015-3542-2},
   number={7},
   journal={The European Physical Journal C},
   publisher={Springer Science and Business Media LLC},
   author={Aad, G. and {\it et. al} {\rm (ATLAS Collaboration)}},
   year={2015},
   month=jul }

@article{Aad_2016,
   title={Measurement of the with the ATLAS Detector},
   volume={116},
   ISSN={1079-7114},
   url={http://dx.doi.org/10.1103/PhysRevLett.116.101801},
   DOI={10.1103/physrevlett.116.101801},
   number={10},
   journal={Physical Review Letters},
   publisher={American Physical Society (APS)},
   author={Aad, G. and {\it et. al} {\rm (ATLAS Collaboration)}},
   year={2016},
   month=mar }

@article{ILC,
       author = {{Baer}, Howard and {Barklow}, Tim and {Fujii}, Keisuke and {Gao}, Yuanning and {Hoang}, Andre and {Kanemura}, Shinya and {List}, Jenny and {Logan}, Heather E. and {Nomerotski}, Andrei and {Perelstein}, Maxim and {Peskin}, Michael E. and {P{\"o}schl}, Roman and {Reuter}, J{\"u}rgen and {Riemann}, Sabine and {Savoy-Navarro}, Aurore and {Servant}, Geraldine and {Tait}, Tim M.~P. and {Yu}, Jaehoon},
        title = "{The International Linear Collider Technical Design Report - Volume 2: Physics}",
      journal = {arXiv e-prints},
         year = 2013,
        month = jun,
          eid = {arXiv:1306.6352},
        pages = {arXiv:1306.6352},
          doi = {10.48550/arXiv.1306.6352},
archivePrefix = {arXiv},
       eprint = {1306.6352},
 primaryClass = {hep-ph},
       adsurl = {https://ui.adsabs.harvard.edu/abs/2013arXiv1306.6352B}
}

@article{CLIC,
    author = "Boland, M J and others",
    editor = "Lebrun, P and Linssen, L and Schulte, D and Sicking, E and Stapnes, S and Thomson, M A and Burrows, P N",
    collaboration = "CLIC, CLICdp",
    title = "{Updated baseline for a staged Compact Linear Collider}",
    eprint = "1608.07537",
    archivePrefix = "arXiv",
    primaryClass = "physics.acc-ph",
    reportNumber = "CERN-2016-004",
    doi = "10.5170/CERN-2016-004",
    journal = "CERN Yellow Rep.",
    volume = "4",
    pages = "1--57",
    year = "2016"
}

@article{ILCHiggs,
author = {Asner, D. and Barklow, T. and Calancha, C. and Fujii, Keisuke and Graf, N. and Haber, Howard and Ishikawa, A. and Kanemura, S. and Kawada, Shin-ichi and Kurata, M. and Miyamoto, Akiya and Neal, H. and Ono, H. and Potter, Cathryn and Strube, Jan and Suehara, T. and Tanabe, T. and Tian, Junping and Tsumura, K. and Yokoya, Hiroshi},
year = {2013},
journal={},
month = {10},
pages = {},
title = {ILC Higgs White Paper}
}

@article{Borzumati_2014,
   title={The Higgs boson and the International Linear Collider},
   volume={2},
   ISSN={2296-424X},
   url={http://dx.doi.org/10.3389/fphy.2014.00032},
   DOI={10.3389/fphy.2014.00032},
   journal={Frontiers in Physics},
   publisher={Frontiers Media SA},
   author={Borzumati, Francesca and Kato, Eriko},
   year={2014},
   month=jun }

@article{Tian:2013yda,
    author = "Tian, Junping and Fujii, Keisuke",
    collaboration = "ILD",
    title = "{Measurement of Higgs couplings and self-coupling at the ILC}",
    eprint = "1311.6528",
    archivePrefix = "arXiv",
    primaryClass = "hep-ph",
    doi = "10.22323/1.180.0316",
    journal = "PoS",
    volume = "EPS-HEP2013",
    pages = "316",
    year = "2013"
}

@article{PhysRevD.23.2001,
  title = {Radiative corrections to Higgs-boson decays in the Weinberg-Salam model},
  author = {Fleischer, J. and Jegerlehner, F.},
  journal = {Phys. Rev. D},
  volume = {23},
  issue = {9},
  pages = {2001--2026},
  numpages = {0},
  year = {1981},
  month = {May},
  publisher = {American Physical Society},
  doi = {10.1103/PhysRevD.23.2001},
  url = {https://link.aps.org/doi/10.1103/PhysRevD.23.2001}
}

@article{KNIEHL19911,
title = {Radiative corrections for H → ZZ in the standard model},
journal = {Nuclear Physics B},
volume = {352},
number = {1},
pages = {1-26},
year = {1991},
issn = {0550-3213},
doi = {https://doi.org/10.1016/0550-3213(91)90126-I},
url = {https://www.sciencedirect.com/science/article/pii/055032139190126I},
author = {Bernd A. Kniehl}
}

@article{Phan:2022amy,
    author = "Phan, Khiem Hong and Tran, Dzung Tri and Nguyen, Anh Thu",
    title = "{One-loop off-shell decay \(H^* \rightarrow ZZ\) at future colliders}",
    eprint = "2209.12410",
    archivePrefix = "arXiv",
    primaryClass = "hep-ph",
    reportNumber = "DTU2022-01",
    doi = "10.15625/0868-3166/18088",
    journal = "Commun. in Phys.",
    volume = "33",
    number = "4",
    pages = "369",
    year = "2023"
}

@article{PhysRevD.107.115031,
  title = {New evaluation of the $HZZ$ coupling: Direct bounds on anomalous contributions and $CP$-violating effects via a new asymmetry},
  author = {Hern\'andez-Ju\'arez, A. I. and Fern\'andez-T\'ellez, A. and Tavares-Velasco, G.},
  journal = {Phys. Rev. D},
  volume = {107},
  issue = {11},
  pages = {115031},
  numpages = {23},
  year = {2023},
  month = {Jun},
  publisher = {American Physical Society},
  doi = {10.1103/PhysRevD.107.115031},
  url = {https://link.aps.org/doi/10.1103/PhysRevD.107.115031}
}

@article{PhysRevD.108.095013,
  title = {Nondecoupling effects from heavy Higgs bosons by matching 2HDM to HEFT amplitudes},
  author = {Arco, F. and Domenech, D. and Herrero, M. J. and Morales, R. A.},
  journal = {Phys. Rev. D},
  volume = {108},
  issue = {9},
  pages = {095013},
  numpages = {29},
  year = {2023},
  month = {Nov},
  publisher = {American Physical Society},
  doi = {10.1103/PhysRevD.108.095013},
  url = {https://link.aps.org/doi/10.1103/PhysRevD.108.095013}
}

@article{KIKUCHI2016807,
title = {Radiative corrections to Higgs coupling constants in two Higgs doublet models},
journal = {Nuclear and Particle Physics Proceedings},
volume = {273-275},
pages = {807-812},
year = {2016},
note = {37th International Conference on High Energy Physics (ICHEP)},
issn = {2405-6014},
doi = {https://doi.org/10.1016/j.nuclphysbps.2015.09.124},
url = {https://www.sciencedirect.com/science/article/pii/S2405601415006136},
author = {Mariko Kikuchi}
}

@article{PhysRevD.96.035014,
  title = {Gauge invariant one-loop corrections to Higgs boson couplings in nonminimal Higgs models},
  author = {Kanemura, Shinya and Kikuchi, Mariko and Sakurai, Kodai and Yagyu, Kei},
  journal = {Phys. Rev. D},
  volume = {96},
  issue = {3},
  pages = {035014},
  numpages = {32},
  year = {2017},
  month = {Aug},
  publisher = {American Physical Society},
  doi = {10.1103/PhysRevD.96.035014},
  url = {https://link.aps.org/doi/10.1103/PhysRevD.96.035014}
}

@article{KANEMURA2016286,
title = {Radiative corrections to the Higgs boson couplings in the model with an additional real singlet scalar field},
journal = {Nuclear Physics B},
volume = {907},
pages = {286-322},
year = {2016},
issn = {0550-3213},
doi = {https://doi.org/10.1016/j.nuclphysb.2016.04.005},
url = {https://www.sciencedirect.com/science/article/pii/S0550321316300402},
author = {Shinya Kanemura and Mariko Kikuchi and Kei Yagyu}
}

@article{Englert:2014ffa,
    author = "Englert, Christoph and Soreq, Yotam and Spannowsky, Michael",
    title = "{Off-Shell Higgs Coupling Measurements in BSM scenarios}",
    eprint = "1410.5440",
    archivePrefix = "arXiv",
    primaryClass = "hep-ph",
    reportNumber = "IPPP-14-91, DCPT-14-182",
    doi = "10.1007/JHEP05(2015)145",
    journal = "JHEP",
    volume = "05",
    pages = "145",
    year = "2015"
}

@article{Baglio_2020,
   title={One-loop corrections to the two-body decays of the neutral Higgs bosons in the complex NMSSM},
   volume={80},
   ISSN={1434-6052},
   url={http://dx.doi.org/10.1140/epjc/s10052-020-08520-8},
   DOI={10.1140/epjc/s10052-020-08520-8},
   number={10},
   journal={The European Physical Journal C},
   publisher={Springer Science and Business Media LLC},
   author={Baglio, Julien and Dao, Thi Nhung and Mühlleitner, Margarete},
   year={2020},
   month=oct }

@article{PhysRevD.87.015012,
  title = {Radiative corrections to the Higgs boson couplings in the triplet model},
  author = {Aoki, Mayumi and Kanemura, Shinya and Kikuchi, Mariko and Yagyu, Kei},
  journal = {Phys. Rev. D},
  volume = {87},
  issue = {1},
  pages = {015012},
  numpages = {39},
  year = {2013},
  month = {Jan},
  publisher = {American Physical Society},
  doi = {10.1103/PhysRevD.87.015012},
  url = {https://link.aps.org/doi/10.1103/PhysRevD.87.015012}
}

@article{Chiang_2017,
   title={Radiative corrections to Higgs couplings with weak gauge bosons in custodial multi-Higgs models},
   volume={774},
   ISSN={0370-2693},
   url={http://dx.doi.org/10.1016/j.physletb.2017.09.061},
   DOI={10.1016/j.physletb.2017.09.061},
   journal={Physics Letters B},
   publisher={Elsevier BV},
   author={Chiang, Cheng-Wei and Kuo, An-Li and Yagyu, Kei},
   year={2017},
   month=nov, pages={119–122} }

@article{Grossman_1994,
   title={Phenomenology of models with more than two Higgs doublets},
   volume={426},
   ISSN={0550-3213},
   url={http://dx.doi.org/10.1016/0550-3213(94)90316-6},
   DOI={10.1016/0550-3213(94)90316-6},
   number={2},
   journal={Nuclear Physics B},
   publisher={Elsevier BV},
   author={Grossman, Yuval},
   year={1994},
   month=sep, pages={355–384} }

@article{Arhrib_2015,
   title={Radiative corrections to the triple Higgs coupling in the inert Higgs doublet model},
   volume={2015},
   ISSN={1029-8479},
   url={http://dx.doi.org/10.1007/JHEP12(2015)007},
   DOI={10.1007/jhep12(2015)007},
   number={12},
   journal={Journal of High Energy Physics},
   publisher={Springer Science and Business Media LLC},
   author={Arhrib, Abdesslam and Benbrik, Rachid and El Falaki, Jaouad and Jueid, Adil},
   year={2015},
   month=dec, pages={1–23} }

@article{PhysRevD.94.115011,
  title = {Testing the dark matter scenario in the inert doublet model by future precision measurements of the Higgs boson couplings},
  author = {Kanemura, Shinya and Kikuchi, Mariko and Sakurai, Kodai},
  journal = {Phys. Rev. D},
  volume = {94},
  issue = {11},
  pages = {115011},
  numpages = {17},
  year = {2016},
  month = {Dec},
  publisher = {American Physical Society},
  doi = {10.1103/PhysRevD.94.115011},
  url = {https://link.aps.org/doi/10.1103/PhysRevD.94.115011}
}

@article{WUDKA,
   title={ELECTROWEAK EFFECTIVE LAGRANGIANS},
   volume={09},
   ISSN={1793-656X},
   url={http://dx.doi.org/10.1142/S0217751X94000959},
   DOI={10.1142/s0217751x94000959},
   number={14},
   journal={International Journal of Modern Physics A},
   publisher={World Scientific Pub Co Pte Lt},
   author={Wudka, José},
   year={1994},
   month=jun, pages={2301–2361} 
   }

@article{Carsten,
  title = {Effective Lagrangians with higher derivatives and equations of motion},
  author = {Knetter, Carsten Grosse},
  journal = {Phys. Rev. D},
  volume = {49},
  issue = {12},
  pages = {6709--6719},
  numpages = {0},
  year = {1994},
  month = {Jun},
  publisher = {American Physical Society},
  doi = {10.1103/PhysRevD.49.6709},
  url = {https://link.aps.org/doi/10.1103/PhysRevD.49.6709}
}

@article{Dobado:1997jx,
    author = "Dobado, A. and Gomez-Nicola, A. and Maroto, Antonio Lopez and Pelaez, J. R.",
    title = "{Effective lagrangians for the standard model}",
    year = "1997",
    journal={}
}

@article{Babu_2001,
   title={Classification of effective neutrino mass operators},
   volume={619},
   ISSN={0550-3213},
   url={http://dx.doi.org/10.1016/S0550-3213(01)00504-1},
   DOI={10.1016/s0550-3213(01)00504-1},
   number={1–3},
   journal={Nuclear Physics B},
   publisher={Elsevier BV},
   author={Babu, K.S. and Leung, C.N.},
   year={2001},
   month=dec, pages={667–689} }

@article{Weinberg,
  title = {Baryon- and Lepton-Nonconserving Processes},
  author = {Weinberg, Steven},
  journal = {Phys. Rev. Lett.},
  volume = {43},
  issue = {21},
  pages = {1566--1570},
  numpages = {0},
  year = {1979},
  month = {Nov},
  publisher = {American Physical Society},
  doi = {10.1103/PhysRevLett.43.1566},
  url = {https://link.aps.org/doi/10.1103/PhysRevLett.43.1566}
}

@article{Leung:1984ni,
    author = "Leung, Chung Ngoc and Love, S. T. and Rao, S.",
    title = "{Low-Energy Manifestations of a New Interaction Scale: Operator Analysis}",
    reportNumber = "FERMILAB-PUB-84-074-T",
    doi = "10.1007/BF01588041",
    journal = "Z. Phys. C",
    volume = "31",
    pages = "433",
    year = "1986"
}

@article{BUCHMULLER1986621,
title = {Effective lagrangian analysis of new interactions and flavour conservation},
journal = {Nuclear Physics B},
volume = {268},
number = {3},
pages = {621-653},
year = {1986},
issn = {0550-3213},
doi = {https://doi.org/10.1016/0550-3213(86)90262-2},
url = {https://www.sciencedirect.com/science/article/pii/0550321386902622},
author = {W. Buchmüller and D. Wyler}
}

@article{Grzadkowski_2010,
   title={Dimension-six terms in the Standard Model Lagrangian},
   volume={2010},
   ISSN={1029-8479},
   url={http://dx.doi.org/10.1007/JHEP10(2010)085},
   DOI={10.1007/jhep10(2010)085},
   number={10},
   journal={Journal of High Energy Physics},
   publisher={Springer Science and Business Media LLC},
   author={Grzadkowski, B. and Iskrzyński, M. and Misiak, M. and Rosiek, J.},
   year={2010},
   month=oct }

@article{LHCHiggs,
    author = "de Florian, D. and others",
    collaboration = "LHC Higgs Cross Section Working Group",
    title = "{Handbook of LHC Higgs Cross Sections: 4. Deciphering the Nature of the Higgs Sector}",
    eprint = "1610.07922",
    archivePrefix = "arXiv",
    primaryClass = "hep-ph",
    reportNumber = "CERN-2017-002-M, CERN-2017-002",
    doi = "10.23731/CYRM-2017-002",
    journal = "CERN Yellow Rep. Monogr.",
    volume = "2",
    pages = "1--869",
    year = "2017"
}

@article{Azatov:2022kbs,
    author = "Azatov, Aleksandr and others",
    title = "{Off-shell Higgs Interpretations Task Force: Models and Effective Field Theories Subgroup Report}",
    eprint = "2203.02418",
    archivePrefix = "arXiv",
    primaryClass = "hep-ph",
    reportNumber = "LHCHWG-2022-001",
    doi = "10.17181/LHCHWG-2022-001",
    month = "3",
    year = "2022",
    journal={}
}

@article{Marzocca2020BSMBF,
  title={BSM Benchmarks for Effective Field Theories in Higgs and Electroweak Physics},
  author={David Marzocca and Francesco Riva and Juan Carlos Criado and Sally Dawson and Jorge de Blas and Brian Quinn Henning and D. Liu and Christopher W. Murphy and M. P{\'e}rez-Victoria and Jos{\'e} Mar{\'i}a Cerver{\'o} Santiago and Luca Vecchi and Lian-tao Wang},
  journal={arXiv: High Energy Physics - Phenomenology},
  year={2020},
  url={https://api.semanticscholar.org/CorpusID:204842371}
}

@article{Falkowski:2015wza,
    author = "Falkowski, Adam and Fuks, Benjamin and Mawatari, Kentarou and Mimasu, Ken and Riva, Francesco and Sanz, Ver{\'o}nica",
    title = "{Rosetta: an operator basis translator for Standard Model effective field theory}",
    eprint = "1508.05895",
    archivePrefix = "arXiv",
    primaryClass = "hep-ph",
    reportNumber = "MCNET-15-25",
    doi = "10.1140/epjc/s10052-015-3806-x",
    journal = "Eur. Phys. J. C",
    volume = "75",
    number = "12",
    pages = "583",
    year = "2015"
}

@article{Falkowski:2001958,
      author        = "Adam Falkowski and Adam Falkowski",
      title         = "{Higgs Basis: Proposal for an EFT basis choice for LHC
                       HXSWG}",
      year          = "2015",
      url           = "https://cds.cern.ch/record/2001958",
      journal={}
}

@article{CAO2025116781,
title = {Unitarity bounds and basis transformations in SMEFT: An analysis of Warsaw and SILH bases},
journal = {Nuclear Physics B},
volume = {1010},
pages = {116781},
year = {2025},
issn = {0550-3213},
doi = {https://doi.org/10.1016/j.nuclphysb.2024.116781},
url = {https://www.sciencedirect.com/science/article/pii/S055032132400347X},
author = {Qing-Hong Cao and Yandong Liu and Shu-Run Yuan}
}

@article{SILH,
   title={The strongly-interacting light Higgs},
   volume={2007},
   ISSN={1029-8479},
   url={http://dx.doi.org/10.1088/1126-6708/2007/06/045},
   DOI={10.1088/1126-6708/2007/06/045},
   number={06},
   journal={Journal of High Energy Physics},
   publisher={Springer Science and Business Media LLC},
   author={Giudice, Gian Francesco and Grojean, Christophe and Pomarol, Alex and Rattazzi, Riccardo},
   year={2007},
   month=jun, pages={045–045} }

@article{SILH1,
    author = "Buchalla, Gerhard and Cata, Oscar and Krause, Claudius",
    title = "{A Systematic Approach to the SILH Lagrangian}",
    eprint = "1412.6356",
    archivePrefix = "arXiv",
    primaryClass = "hep-ph",
    reportNumber = "LMU-ASC-49-14, FLAVOUR(267104)-ERC-77",
    doi = "10.1016/j.nuclphysb.2015.03.024",
    journal = "Nucl. Phys. B",
    volume = "894",
    pages = "602--620",
    year = "2015"
}

@article{SILH2,
    author = "Alloul, Adam and Fuks, Benjamin and Sanz, Ver{\'o}nica",
    title = "{Phenomenology of the Higgs Effective Lagrangian via FEYNRULES}",
    eprint = "1310.5150",
    archivePrefix = "arXiv",
    primaryClass = "hep-ph",
    reportNumber = "CERN-PH-TH-2013-248",
    doi = "10.1007/JHEP04(2014)110",
    journal = "JHEP",
    volume = "04",
    pages = "110",
    year = "2014"
}

@article{Pilaftsis,
   title={Radiatively induced neutrino masses and large Higgs-neutrino couplings in the Standard Model with Majorana fields},
   volume={55},
   ISSN={1434-6052},
   url={http://dx.doi.org/10.1007/BF01482590},
   DOI={10.1007/bf01482590},
   number={2},
   journal={Zeitschrift f\"ur Physik C Particles and Fields},
   publisher={Springer Science and Business Media LLC},
   author={Pilaftsis, Apostolos},
   year={1992},
   month=jun, pages={275–282} }

@article{WWA,
    author = "Mart{\'\i}nez, Eduardo and Monta{\~n}o-Dom{\'\i}nguez, Javier and Novales-S{\'a}nchez, H{\'e}ctor and Salinas, M{\'o}nica",
    title = "{New physics in WW{\ensuremath{\gamma}} at one loop via Majorana neutrinos}",
    eprint = "2211.04629",
    archivePrefix = "arXiv",
    primaryClass = "hep-ph",
    doi = "10.1103/PhysRevD.107.035025",
    journal = "Phys. Rev. D",
    volume = "107",
    number = "3",
    pages = "035025",
    year = "2023"
}

@article{ZZZ,
    author = "Novales-S{\'a}nchez, H{\'e}ctor and Salinas, M{\'o}nica",
    title = "{Majorana neutrinos in the triple gauge boson coupling ZZZ*}",
    eprint = "2309.02400",
    archivePrefix = "arXiv",
    primaryClass = "hep-ph",
    doi = "10.1103/PhysRevD.108.075032",
    journal = "Phys. Rev. D",
    volume = "108",
    number = "7",
    pages = "075032",
    year = "2023"
}

@article{WWZ,
    author = "Novales-S{\'a}nchez, H{\'e}ctor and Salinas, M{\'o}nica and V{\'a}zquez-Castro, Humberto",
    title = "{One-loop contributions to WWZ from a seesaw variant with radiatively induced light neutrino masses}",
    eprint = "2404.08205",
    archivePrefix = "arXiv",
    primaryClass = "hep-ph",
    doi = "10.1103/PhysRevD.110.035025",
    journal = "Phys. Rev. D",
    volume = "110",
    number = "3",
    pages = "035025",
    year = "2024"
}

@article{CLFV,
    author = "Ram{\'\i}rez, Enrique and Novales-S{\'a}nchez, H{\'e}ctor and V{\'a}zquez-Castro, Humberto and Salinas, M{\'o}nica",
    title = "{Effects of virtual Majorana neutrinos on charged lepton flavor violation decays from a seesaw variant with radiatively induced light neutrino masses}",
    eprint = "2505.09051",
    archivePrefix = "arXiv",
    primaryClass = "hep-ph",
    doi = "10.1088/1361-6471/ae0c24",
    journal = "J. Phys. G",
    volume = "52",
    number = "10",
    pages = "105004",
    year = "2025"
}

@article{WWh,
    author = "Novales-S{\'a}nchez, H{\'e}ctor and Ram{\'\i}rez, Enrique and Salinas, M{\'o}nica and V{\'a}zquez-Castro, Humberto",
    title = "{WWh anomalous couplings at one loop from a seesaw variant of radiatively-induced neutrino masses}",
    eprint = "2507.05574",
    archivePrefix = "arXiv",
    primaryClass = "hep-ph",
    doi = "10.1007/JHEP11(2025)009",
    journal = "JHEP",
    volume = "11",
    pages = "009",
    year = "2025"
}

@article{DMneutrinos,
  title = {Space-Time Approach to Quantum Electrodynamics},
  author = {Feynman, R. P.},
  journal = {Phys. Rev.},
  volume = {76},
  issue = {6},
  pages = {769--789},
  numpages = {0},
  year = {1949},
  month = {Sep},
  publisher = {American Physical Society},
  doi = {10.1103/PhysRev.76.769},
  url = {https://link.aps.org/doi/10.1103/PhysRev.76.769}
}

@article{DMneutrinos2,
  title = {Feynman rules for Majorana-neutrino interactions},
  author = {Gluza, J. and Zraek, M.},
  journal = {Phys. Rev. D},
  volume = {45},
  issue = {5},
  pages = {1693--1700},
  numpages = {0},
  year = {1992},
  month = {Mar},
  publisher = {American Physical Society},
  doi = {10.1103/PhysRevD.45.1693},
  url = {https://link.aps.org/doi/10.1103/PhysRevD.45.1693}
}

@article{Denner:1992me,
    author = "Denner, Ansgar and Eck, H. and Hahn, O. and Kublbeck, J.",
    title = "{Compact Feynman rules for Majorana fermions}",
    reportNumber = "PRINT-92-0315 (CERN)",
    doi = "10.1016/0370-2693(92)91045-B",
    journal = "Phys. Lett. B",
    volume = "291",
    pages = "278--280",
    year = "1992"
}

@article{Gates:1987ay,
    author = "Gates, Evalyn I. and Kowalski, Kenneth L.",
    title = "{MAJORANA FEYNMAN RULES}",
    reportNumber = "ANL-HEP-PR-87-55",
    doi = "10.1103/PhysRevD.37.938",
    journal = "Phys. Rev. D",
    volume = "37",
    pages = "938",
    year = "1988"
}

@article{Haber:1984rc,
    author = "Haber, Howard E. and Kane, Gordon L.",
    title = "{The Search for Supersymmetry: Probing Physics Beyond the Standard Model}",
    reportNumber = "UM-HE-TH-83-17, SCIPP-85-47",
    doi = "10.1016/0370-1573(85)90051-1",
    journal = "Phys. Rept.",
    volume = "117",
    pages = "75--263",
    year = "1985"
}

@article{Wick,
  title = {The Evaluation of the Collision Matrix},
  author = {Wick, G. C.},
  journal = {Phys. Rev.},
  volume = {80},
  issue = {2},
  pages = {268--272},
  numpages = {0},
  year = {1950},
  month = {Oct},
  publisher = {American Physical Society},
  doi = {10.1103/PhysRev.80.268},
  url = {https://link.aps.org/doi/10.1103/PhysRev.80.268}
}

@article{DENNER,
title = {Feynman rules for fermion-number-violating interactions},
journal = {Nuclear Physics B},
volume = {387},
number = {2},
pages = {467-481},
year = {1992},
issn = {0550-3213},
doi = {https://doi.org/10.1016/0550-3213(92)90169-C},
url = {https://www.sciencedirect.com/science/article/pii/055032139290169C},
author = {A. Denner and H. Eck and O. Hahn and J. Küblbeck}
}

@article{Bollini,
    author = "Bollini, C. G. and Giambiagi, J. J.",
    title = "{Dimensional Renormalization: The Number of Dimensions as a Regularizing Parameter}",
    doi = "10.1007/BF02895558",
    journal = "Nuovo Cim. B",
    volume = "12",
    pages = "20--26",
    year = "1972"
}

@article{THOOFT,
title = {Regularization and renormalization of gauge fields},
journal = {Nuclear Physics B},
volume = {44},
number = {1},
pages = {189-213},
year = {1972},
issn = {0550-3213},
doi = {https://doi.org/10.1016/0550-3213(72)90279-9},
url = {https://www.sciencedirect.com/science/article/pii/0550321372902799},
author = {G. {'t Hooft} and M. Veltman}
}

@article{PASSARINO1979151,
title = {One-loop corrections for e+e− annihilation into μ+μ− in the Weinberg model},
journal = {Nuclear Physics B},
volume = {160},
number = {1},
pages = {151-207},
year = {1979},
issn = {0550-3213},
doi = {https://doi.org/10.1016/0550-3213(79)90234-7},
url = {https://www.sciencedirect.com/science/article/pii/0550321379902347},
author = {G. Passarino and M. Veltman}
}

@article{DEVARAJ1998483,
title = {Reduction of one-loop tensor form factors to scalar integrals: a general scheme},
journal = {Nuclear Physics B},
volume = {519},
number = {1},
pages = {483-513},
year = {1998},
issn = {0550-3213},
doi = {https://doi.org/10.1016/S0550-3213(98)00035-2},
url = {https://www.sciencedirect.com/science/article/pii/S0550321398000352},
author = {Ganesh Devaraj and Robin G. Stuart}
}

@article{FeynCalc1,
title = {Feyn Calc - Computer-algebraic calculation of Feynman amplitudes},
journal = {Computer Physics Communications},
volume = {64},
number = {3},
pages = {345-359},
year = {1991},
issn = {0010-4655},
doi = {https://doi.org/10.1016/0010-4655(91)90130-D},
url = {https://www.sciencedirect.com/science/article/pii/001046559190130D},
author = {R. Mertig and M. Böhm and A. Denner}
}

@article{FeynCalc2,
   title={New developments in FeynCalc 9.0},
   volume={207},
   ISSN={0010-4655},
   url={http://dx.doi.org/10.1016/j.cpc.2016.06.008},
   DOI={10.1016/j.cpc.2016.06.008},
   journal={Computer Physics Communications},
   publisher={Elsevier BV},
   author={Shtabovenko, Vladyslav and Mertig, Rolf and Orellana, Frederik},
   year={2016},
   month=oct, pages={432–444} 
}

@article{FeynCalc3,
   title={FeynCalc 9.3: New features and improvements},
   volume={256},
   ISSN={0010-4655},
   url={http://dx.doi.org/10.1016/j.cpc.2020.107478},
   DOI={10.1016/j.cpc.2020.107478},
   journal={Computer Physics Communications},
   publisher={Elsevier BV},
   author={Shtabovenko, Vladyslav and Mertig, Rolf and Orellana, Frederik},
   year={2020},
   month=nov, pages={107478} 
}

@article{Patel_2015,
   title={Package-X: A Mathematica package for the analytic calculation of one-loop integrals},
   volume={197},
   ISSN={0010-4655},
   url={http://dx.doi.org/10.1016/j.cpc.2015.08.017},
   DOI={10.1016/j.cpc.2015.08.017},
   journal={Computer Physics Communications},
   publisher={Elsevier BV},
   author={Patel, Hiren H.},
   year={2015},
   month=dec, pages={276–290}
}

@article{Alam,
  title = {Completed SDSS-IV extended Baryon Oscillation Spectroscopic Survey: Cosmological implications from two decades of spectroscopic surveys at the Apache Point Observatory},
  author = {Alam, Shadab and \it et. al \rm(eBOSS)},
  journal = {Phys. Rev. D},
  volume = {103},
  issue = {8},
  pages = {083533},
  numpages = {43},
  year = {2021},
  month = {Apr},
  publisher = {American Physical Society},
  doi = {10.1103/PhysRevD.103.083533},
  url = {https://link.aps.org/doi/10.1103/PhysRevD.103.083533}
}

@article{ refId0,
	author = {{Aghanim, N.} and {\it et. al \rm(Plank)}},
	title = {Planck 2018 results - VI. Cosmological parameters},
	DOI= "10.1051/0004-6361/201833910",
	url= "https://doi.org/10.1051/0004-6361/201833910",
	journal = {A\&A},
	year = 2020,
	volume = 641,
	pages = "A6",
}

@article{Katrin,
   title={Direct neutrino-mass measurement based on 259 days of KATRIN data},
   volume={388},
   ISSN={1095-9203},
   url={http://dx.doi.org/10.1126/science.adq9592},
   DOI={10.1126/science.adq9592},
   number={6743},
   journal={Science},
   publisher={American Association for the Advancement of Science (AAAS)},
   author={Aker, Max and \it et. al \rm(KATRIN)},
   year={2025},
   month=apr, pages={180–185} }

@article{Abe_2024,
   title={Solar neutrino measurements using the full data period of Super-Kamiokande-IV},
   volume={109},
   ISSN={2470-0029},
   url={http://dx.doi.org/10.1103/PhysRevD.109.092001},
   DOI={10.1103/physrevd.109.092001},
   number={9},
   journal={Physical Review D},
   publisher={American Physical Society (APS)},
   author={Abe, K. and \it et. al \rm(Super-Kamiokand)},
   year={2024},
   month=may }

@article{Minos,
   title={Precision Constraints for Three-Flavor Neutrino Oscillations from the Full 
<mml:math xmlns:mml="http://www.w3.org/1998/Math/MathML" display="inline"><mml:mrow><mml:mi>MINOS</mml:mi><mml:mo>+</mml:mo></mml:mrow></mml:math>
 and MINOS Dataset},
   volume={125},
   ISSN={1079-7114},
   url={http://dx.doi.org/10.1103/PhysRevLett.125.131802},
   DOI={10.1103/physrevlett.125.131802},
   number={13},
   journal={Physical Review Letters},
   publisher={American Physical Society (APS)},
   author={Adamson, P. and \it et. al \rm(MINOS+)},
   year={2020},
   month=sep 
}

@article{Nova,
   title={Improved measurement of neutrino oscillation parameters by the NOvA experiment},
   volume={106},
   ISSN={2470-0029},
   url={http://dx.doi.org/10.1103/PhysRevD.106.032004},
   DOI={10.1103/physrevd.106.032004},
   number={3},
   journal={Physical Review D},
   publisher={American Physical Society (APS)},
   author={Acero, M. A. and \it et. al \rm(NOvA)},
   year={2022},
   month=aug }

@article{T2k,
   title={Measurements of neutrino oscillation parameters from the T2K experiment using $$3.6\times 10^{21}$$ protons on target},
   volume={83},
   ISSN={1434-6052},
   url={http://dx.doi.org/10.1140/epjc/s10052-023-11819-x},
   DOI={10.1140/epjc/s10052-023-11819-x},
   number={9},
   journal={The European Physical Journal C},
   publisher={Springer Science and Business Media LLC},
   author={Abe, K. and \it et. al \rm(T2K)},
   year={2023},
   month=sep }

@article{Icecube,
   title={Measurement of atmospheric neutrino mixing with improved IceCube DeepCore calibration and data processing},
   volume={108},
   ISSN={2470-0029},
   url={http://dx.doi.org/10.1103/PhysRevD.108.012014},
   DOI={10.1103/physrevd.108.012014},
   number={1},
   journal={Physical Review D},
   publisher={American Physical Society (APS)},
   author={Abbasi, R. and \it et. al \rm(IceCube Collaboration)},
   year={2023},
   month=jul }

@article{SK,
   title={Atmospheric neutrino oscillation analysis with neutron tagging and an expanded fiducial volume in Super-Kamiokande I–V},
   volume={109},
   ISSN={2470-0029},
   url={http://dx.doi.org/10.1103/PhysRevD.109.072014},
   DOI={10.1103/physrevd.109.072014},
   number={7},
   journal={Physical Review D},
   publisher={American Physical Society (APS)},
   author={Wester, T. and \it et. al \rm(Super-Kamiokande)},
   year={2024},
   month=apr }

@article{km3net,
      title={Measurement of neutrino oscillation parameters with the first six detection units of KM3NeT/ORCA}, 
      author={S. Aiello and \it et. al \rm(KM3NeT) },
      year={2024},
      eprint={2408.07015},
      archivePrefix={arXiv},
      primaryClass={hep-ex},
      url={https://arxiv.org/abs/2408.07015},
      journal={},
}

@article{Reno,
  title = {Measurement of Reactor Antineutrino Oscillation Amplitude and Frequency at RENO},
  author = {Bak, G. and \it et. al \rm(RENO)},
  journal = {Phys. Rev. Lett.},
  volume = {121},
  issue = {20},
  pages = {201801},
  numpages = {6},
  year = {2018},
  month = {Nov},
  publisher = {American Physical Society},
  doi = {10.1103/PhysRevLett.121.201801},
  url = {https://link.aps.org/doi/10.1103/PhysRevLett.121.201801}
}

@article{DayaBay,
   title={Measurement of Electron Antineutrino Oscillation Amplitude and Frequency via Neutron Capture on Hydrogen at Daya Bay},
   volume={133},
   ISSN={1079-7114},
   url={http://dx.doi.org/10.1103/PhysRevLett.133.151801},
   DOI={10.1103/physrevlett.133.151801},
   number={15},
   journal={Physical Review Letters},
   publisher={American Physical Society (APS)},
   author={An, F. P. and \it et. al \rm(Daya Bay)},
   year={2024},
   month=oct }

@article{Nova2,
   title={Expanding neutrino oscillation parameter measurements in NOvA using a Bayesian approach},
   volume={110},
   ISSN={2470-0029},
   url={http://dx.doi.org/10.1103/PhysRevD.110.012005},
   DOI={10.1103/physrevd.110.012005},
   number={1},
   journal={Physical Review D},
   publisher={American Physical Society (APS)},
   author={Acero, M. A. and \it et. al \rm(NovA) },
   year={2024},
   month=jul }

@article{dayabay2,
      title={Precision measurement of reactor antineutrino oscillation at kilometer-scale baselines by Daya Bay}, 
      author={F. P. An and \it et. al \rm(Daya Bay)},
      year={2022},
      eprint={2211.14988},
      archivePrefix={arXiv},
      primaryClass={hep-ex},
      url={https://arxiv.org/abs/2211.14988},
      journal={}
}

@article{PhysRevLett.120.221801,
  title = {Search for Heavy Neutral Leptons in Events with Three Charged Leptons in Proton-Proton Collisions at $\sqrt{s}=13\text{ }\text{ }\mathrm{TeV}$},
  author = {Sirunyan, A. M. and \it et. al \rm (CMS)},
  journal = {Phys. Rev. Lett.},
  volume = {120},
  issue = {22},
  pages = {221801},
  numpages = {20},
  year = {2018},
  month = {May},
  publisher = {American Physical Society},
  doi = {10.1103/PhysRevLett.120.221801},
  url = {https://link.aps.org/doi/10.1103/PhysRevLett.120.221801}
}

@article{Ellis_2018,
   title={Updated global SMEFT fit to Higgs, diboson and electroweak data},
   volume={2018},
   ISSN={1029-8479},
   url={http://dx.doi.org/10.1007/JHEP06(2018)146},
   DOI={10.1007/jhep06(2018)146},
   number={6},
   journal={Journal of High Energy Physics},
   publisher={Springer Science and Business Media LLC},
   author={Ellis, John and Murphy, Christopher W. and Sanz, Verónica and You, Tevong},
   year={2018},
   month=jun }

@article{Denizli_2018,
   title={Constraints on Higgs Effective Couplings in <mml:math xmlns:mml="http://www.w3.org/1998/Math/MathML" id="M1"><mml:mi>H</mml:mi><mml:mi>ν</mml:mi><mml:mover accent="false"><mml:mrow><mml:mi>ν</mml:mi></mml:mrow><mml:mo>¯</mml:mo></mml:mover></mml:math> Production of CLIC at 380 GeV},
   volume={2018},
   ISSN={1687-7365},
   url={http://dx.doi.org/10.1155/2018/1627051},
   DOI={10.1155/2018/1627051},
   journal={Advances in High Energy Physics},
   publisher={Wiley},
   author={Denizli, H. and Senol, A.},
   year={2018},
   pages={1–8} }

@article{ATLAS-CONF-2019-029,
    author = "ATLAS Collaboration",
    title = "{Measurements and interpretations of Higgs-boson fiducial cross sections in the diphoton decay channel using 139 fb$^−1$ of $pp$ collision data at $\sqrt{s}$ = 13 TeV with the ATLAS detector}",
    journal = "ATLAS-CONF-2019-029",
    month = "7",
    year = "2019"
}

@article{Karadeniz_2020,
   title={CP-violating Higgs-gauge boson couplings in $$H\nu \bar{\nu }$$ production at three energy stages of CLIC},
   volume={80},
   ISSN={1434-6052},
   url={http://dx.doi.org/10.1140/epjc/s10052-020-7740-1},
   DOI={10.1140/epjc/s10052-020-7740-1},
   number={3},
   journal={The European Physical Journal C},
   publisher={Springer Science and Business Media LLC},
   author={Karadeniz, O. and Senol, A. and Oyulmaz, K. Y. and Denizli, H.},
   year={2020},
   month=mar }

@article{Alam_2017,
   title={Constraining capability of Zγh production at the ILC},
   volume={32},
   ISSN={1793-656X},
   url={http://dx.doi.org/10.1142/S0217751X17500178},
   DOI={10.1142/s0217751x17500178},
   number={02n03},
   journal={International Journal of Modern Physics A},
   publisher={World Scientific Pub Co Pte Lt},
   author={Alam, Sher and Behera, Subhasish and Kumar, Satendra and Sahoo, Shibananda},
   year={2017},
   month=jan, pages={1750017} }

@article{Spor_2025,
   title={Constraints on the anomalous Higgs boson couplings in
                    <mml:math xmlns:mml="http://www.w3.org/1998/Math/MathML" altimg="si37.svg">
                      <mml:mrow>
                        <mml:mi>Z</mml:mi>
                        <mml:mi>γ</mml:mi>
                        <mml:mi>γ</mml:mi>
                      </mml:mrow>
                    </mml:math>
                    production at muon collider},
   volume={1020},
   ISSN={0550-3213},
   url={http://dx.doi.org/10.1016/j.nuclphysb.2025.117170},
   DOI={10.1016/j.nuclphysb.2025.117170},
   journal={Nuclear Physics B},
   publisher={Elsevier BV},
   author={Spor, Serdar},
   year={2025},
   month=nov, pages={117170} }

@article{Gurkanli_2025,
   title={Probing CP-violating Higgs–Gauge Boson Couplings at the Future Muon Collider},
   volume={2025},
   ISSN={2050-3911},
   url={http://dx.doi.org/10.1093/ptep/ptaf068},
   DOI={10.1093/ptep/ptaf068},
   number={5},
   journal={Progress of Theoretical and Experimental Physics},
   publisher={Oxford University Press (OUP)},
   author={Gurkanli, Emre and Spor, Serdar},
   year={2025},
   month=may }

@article{sym13071256,
author = {Arbuzov, Andrej and Bondarenko, Serge and Kalinovskaya, Lidia and Sadykov, Renat and Yermolchyk, Vitaly},
title = {Electroweak Effects in e+e−→ZH Process},
journal = {Symmetry},
volume = {13},
year = {2021},
number = {7},
DOI = {10.3390/sym13071256}
}

@article{Jegerlehner_2005,
   title={One-loop electroweak factorizable correctionsfor the Higgsstrahlung at a linear collider},
   volume={44},
   ISSN={1434-6052},
   url={http://dx.doi.org/10.1140/epjc/s2005-02363-1},
   DOI={10.1140/epjc/s2005-02363-1},
   number={2},
   journal={The European Physical Journal C},
   publisher={Springer Science and Business Media LLC},
   author={Jegerlehner, F. and Kołodziej, K. and Westwański, T.},
   year={2005},
   month=oct, pages={195–203} }

@article{Bondarenko,
  title = {One-loop electroweak radiative corrections to polarized ${e}^{+}{e}^{\ensuremath{-}}\ensuremath{\rightarrow}ZH$},
  author = {Bondarenko, S. and Dydyshka, Ya. and Kalinovskaya, L. and Rumyantsev, L. and Sadykov, R. and Yermolchyk, V.},
  journal = {Phys. Rev. D},
  volume = {100},
  issue = {7},
  pages = {073002},
  numpages = {6},
  year = {2019},
  month = {Oct},
  publisher = {American Physical Society},
  doi = {10.1103/PhysRevD.100.073002},
  url = {https://link.aps.org/doi/10.1103/PhysRevD.100.073002}
}

@article{BELANGER2003252,
title = {Full one-loop electroweak radiative corrections to single Higgs production in e+e−},
journal = {Physics Letters B},
volume = {559},
number = {3},
pages = {252-262},
year = {2003},
issn = {0370-2693},
doi = {https://doi.org/10.1016/S0370-2693(03)00339-3},
url = {https://www.sciencedirect.com/science/article/pii/S0370269303003393},
author = {G Bélanger and F Boudjema and J Fujimoto and T Ishikawa and T Kaneko and K Kato and Y Shimizu}
}

@article{Freitas,
  title = {Two-Loop Electroweak Corrections with Fermion Loops to ${e}^{+}{e}^{\ensuremath{-}}\ensuremath{\rightarrow}ZH$},
  author = {Freitas, Ayres and Song, Qian},
  journal = {Phys. Rev. Lett.},
  volume = {130},
  issue = {3},
  pages = {031801},
  numpages = {6},
  year = {2023},
  month = {Jan},
  publisher = {American Physical Society},
  doi = {10.1103/PhysRevLett.130.031801},
  url = {https://link.aps.org/doi/10.1103/PhysRevLett.130.031801}
}

@article{Liu_2014,
   title={Full one-loop electroweak corrections to e + e − → ZHγ at a Higgs factory},
   volume={2014},
   ISSN={1029-8479},
   url={http://dx.doi.org/10.1007/JHEP04(2014)189},
   DOI={10.1007/jhep04(2014)189},
   number={4},
   journal={Journal of High Energy Physics},
   publisher={Springer Science and Business Media LLC},
   author={Liu, Ning and Ren, Jie and Wu, Lei and Wu, Peiwen and Yang, Jin Min},
   year={2014},
   month=apr }

@article{Higgs:1964pj,
    author = "Higgs, Peter W.",
    editor = "Taylor, J. C.",
    title = "{Broken Symmetries and the Masses of Gauge Bosons}",
    doi = "10.1103/PhysRevLett.13.508",
    journal = "Phys. Rev. Lett.",
    volume = "13",
    pages = "508--509",
    year = "1964"
}

@article{Englert:1964et,
    author = "Englert, F. and Brout, R.",
    editor = "Taylor, J. C.",
    title = "{Broken Symmetry and the Mass of Gauge Vector Mesons}",
    doi = "10.1103/PhysRevLett.13.321",
    journal = "Phys. Rev. Lett.",
    volume = "13",
    pages = "321--323",
    year = "1964"
}

@article{DeBlas:2019qco,
    author = "De Blas, Jorge and Durieux, Gauthier and Grojean, Christophe and Gu, Jiayin and Paul, Ayan",
    title = "{On the future of Higgs, electroweak and diboson measurements at lepton colliders}",
    eprint = "1907.04311",
    archivePrefix = "arXiv",
    primaryClass = "hep-ph",
    reportNumber = "DESY-19-077, DESY 19-077, HU-EP-19/33, MITP/19-028",
    doi = "10.1007/JHEP12(2019)117",
    journal = "JHEP",
    volume = "12",
    pages = "117",
    year = "2019"
}

@article{Sha:2022bkt,
    author = "Sha, Qiyu and others",
    title = "{Probing Higgs CP properties at the CEPC in the $e^{+} e^{-} \rightarrow Z H \rightarrow l^{+} l^{-}H$ using optimal variables}",
    eprint = "2203.11707",
    archivePrefix = "arXiv",
    primaryClass = "hep-ex",
    doi = "10.1140/epjc/s10052-022-10926-5",
    journal = "Eur. Phys. J. C",
    volume = "82",
    number = "11",
    pages = "981",
    year = "2022",
    note = "[Erratum: Eur.Phys.J.C 83, 62 (2023)]"
}

@article{Teiji:TAKAGI192483,
  title={On an Algebraic Problem Reluted to an Analytic Theorem of Carath&eacute;odory and Fej&eacute;r and on an Allied Theorem of Landau},
  author={Teiji Takagi},
  journal={Jpn. J. Math.},
  volume={1},
  number={ },
  pages={83-93},
  year={1924},
  doi={10.4099/jjm1924.1.0_83}
}

%

\end{document}